\documentclass{aa}  

\usepackage{graphicx}
\usepackage{txfonts}
\usepackage{lipsum}
\usepackage{subcaption}         
\usepackage{lscape}             
\usepackage{placeins}           
                                
\usepackage[colorlinks=true,linkcolor=blue,citecolor=blue,urlcolor=blue]{hyperref}
\usepackage{multirow}

\begin{document}

\title{Nebular continuum in high-redshift galaxies with JWST}

\author{Henrique Miranda\inst{1,2}\fnmsep\thanks{Corresponding author: \href{mailto:hbmiranda@fc.ul.pt}{hbmiranda@fc.ul.pt}}
       \and Ciro Pappalardo\inst{1,2}
       \and José Afonso\inst{1,2}
       \and Polychronis Papaderos\inst{1,3}
       \and Rodrigo Carvajal\inst{1}
       }

\institute{Instituto de Astrof\'{i}sica e Ci\^{e}ncias do Espa\c{c}o, Universidade de Lisboa - OAL, Tapada da Ajuda, PT1349-018 Lisboa, Portugal
          \and
          Departamento de F\'{i}sica, Faculdade de Ci\^{e}ncias da Universidade de Lisboa, Edif\'{i}cio C8, Campo Grande, PT1749-016 Lisboa, Portugal
          \and
          Instituto de Astrofísica e Ciências do Espaço, Universidade do Porto – CAUP, Rua das Estrelas, PT4150-762 Porto, Portugal
          }
          

\abstract
{Studying the relevance of accounting for both stellar and nebular continuum emission when performing optical spectral fitting has been mainly limited to galaxies in the local Universe. Since the start of operations of the \textit{James Webb} Space Telescope (JWST), a wealth of high-quality spectroscopic data has been gathered for high-redshift galaxies, opening the possibility of carrying out these studies in the younger Universe.}
{We aim to estimate the nebular continuum contribution (X$_{\text{neb}}$) of high-redshift star-forming galaxies, and correlate it with different established tracers and physical properties. Furthermore, we test the previously established threshold for significant nebular contribution (X$_{\text{neb}}$ $=$ 8\%) and characterise the physical and evolutionary properties of galaxies above and below this limit.}
{We select a sample of 54 star-forming galaxies from the DAWN JWST Archive meeting quality criteria required to perform optical spectral fitting with the population spectral synthesis code FADO, namely R $\sim$ 1000 and median signal-to-noise ratio per pixel above 5. The selected galaxies cover the redshift range 1.7 $<$ $z$ $<$ 3.9 and we fit their rest-frame optical spectra to estimate the stellar and nebular continuum emission, plus the physical and evolutionary properties.}
{We show that the H$\alpha$ and H$\beta$ equivalent widths remain suitable tracers of X$_{\text{neb}}$ at high redshift. Additionally, our results are consistent with the threshold of X$_{\text{neb}}$ $=$ 8\%, above which the nebular continuum contribution significantly affects the modelling of galaxies when using purely stellar population spectral synthesis codes. We find that galaxies above this threshold are generally less massive, younger, exhibit higher star-forming activity and lower dust extinction relative to those below the threshold. Finally, the results suggest that galaxies with X$_{\text{neb}}$ $>$ 8\% are dominated in both light and mass by stellar populations younger than 20 Myr, although the inferred mass fraction of these young stellar populations is likely overestimated due to the outshining effect.}
{Considering the rapidly growing amount of high-quality JWST spectroscopic data and the arrival of future instruments targeting the peak epoch of star formation in the Universe, such as MOONS, this work underscores the importance of properly accounting for both the stellar and nebular continuum emission when performing optical spectral fitting of high-redshift star-forming galaxies. Accurately treating these components is essential for a reliable estimation of their stellar populations and physical properties, thus improving our understanding of galaxy evolution.}


\maketitle
\nolinenumbers

\section{Introduction}

Understanding the processes that regulate how galaxies assemble their mass is a fundamental goal of extragalactic astrophysics. In this context, galaxies with substantial star-forming (SF) activity are especially important, as they are undergoing a phase of their evolution for which there is a high rate of stellar mass growth. These galaxies, also called extreme emission-line galaxies (EELGs), are a diverse population with different subgroups with distinct characteristics, such as blue compact dwarfs, HII galaxies and green peas. These objects have been extensively characterised from low  to high redshift \citep[e.g.][]{salzer89,terlevich91,guseva01,guseva07,cardamone09,amorin10,amorin15,atek11,vanderwel11,brunker20,boyett22,llerena24}.

One fundamental aspect of these systems is their high nebular continuum emission at optical wavelengths, which arises from the ionisation of the nebular gas surrounding young stellar populations. In some cases, this emission is non-negligible relative to the stellar emission, and properly considering both stellar and nebular components is fundamental for accurately retrieving the physical and evolutionary properties of galaxies \citep[e.g.][]{krueger95,izotov97,izotov11,papaderos98,papaderos02,anders03,schaerer09,schaerer10,cardoso19,cardoso22,pappalardo21,breda22,miranda23,miranda25}.

In this regard, the population spectral synthesis code FADO \citep[Fitting Analysis using Differential evolution Optimization;][]{gomes17}, was specifically designed to handle self-consistently both the stellar and nebular continuum emission when fitting the optical spectrum of galaxies. This tool has been used to understand the relevance of the nebular continuum emission in optical spectral fitting and to accurately characterise the physical properties of SF galaxies in the local Universe \citep{cardoso19,cardoso22,pappalardo21,breda22,miranda23,miranda25}. In particular, \cite{miranda25} correlated the nebular contribution, i.e. the ratio between the nebular and total continuum in the optical (X$_{\text{neb}}$), with the equivalent width (EW) of the H$\alpha$ and H$\beta$ emission lines. Their results also showed that the self-consistent consideration of the stellar and nebular continuum emission is fundamental to accurately estimate the physical properties of galaxies when X$_{\text{neb}}$ $\geq$ 8\%, corresponding to EW(H$\alpha$) $\geq$ 500 \AA.

These studies require high-quality spectroscopic data, namely a sufficiently high signal-to-noise ratio (S/N) in the continuum and good spectral resolution, plus a significant coverage of the optical rest-frame. Additionally, EELGs are typically metal-poor, and thus tend to have intrinsically low luminosities (as implied by the luminosity-metallicity relation by \citealt{guseva09}), and are therefore difficult to detect at high redshifts. Due to these facts, these studies have so far been largely confined to the local Universe, particularly using SDSS \citep[Sloan Digital Sky Survey;][]{york00}. Although there are spectroscopic surveys of high-redshift galaxies (e.g. zCOSMOS \citep{lilly07}, MOSDEF \citep{kriek15}, VANDELS \citep{mclure18,pentericci18}), none of them meet all these criteria simultaneously for a significant number of objects. However, since the \textit{James Webb} Space Telescope (JWST) began operating, thousands of high-quality spectra of high-redshift galaxies have been collected, providing the possibility of studying the importance of the nebular continuum emission in the early Universe.

Evidence shows that the SF activity of the Universe was greater in the past, as indicated by the evolution with redshift of the star-forming main sequence \citep[SFMS; e.g.][]{speagle14}, the star formation rate density \citep[e.g.][]{madau14}, and the EW of H$\alpha$ \citep[e.g.][]{faisst16}. Thus, while the nebular continuum emission is only significant in extreme cases within the local Universe \citep[e.g.][]{izotov11,cardoso22,miranda23}, it is expected to be a more widespread phenomenon at higher redshifts \citep{miranda25,katz25,trussler25}. In fact, recent JWST observations have revealed an increase in the fraction of EELGs at higher redshifts \citep{boyett24} and even enabled the detection of potential nebular-dominated galaxies \citep{cameron24}. All in all, understanding the relevance of properly accounting for the nebular continuum emission in the optical is even more fundamental at high redshifts.

Leveraging the high-quality spectroscopic data from JWST, we aim to gather a sample of galaxy spectra that meet the necessary quality criteria to carry out full spectral fitting and study the nebular contribution at high redshifts. In this work, we expand to higher-redshifts the study of the previously established relation between X$_{\text{neb}}$ and different tracers and also the threshold to identify galaxies with significant nebular contribution. Additionally, we characterise and compare the physical and evolutionary properties of galaxies with and without significant nebular contribution at both low and high redshifts.

The paper is organised as follows. Section \ref{sec:sample_methods} describes the sample selection and the spectral fitting methodology. In Sect. \ref{sec:FitQuality_NebCont}, we analyse the quality of the spectral fitting results, estimate the nebular contribution and relate it to relevant tracers and physical properties of galaxies. In Sect. \ref{sec:HEW_galaxies_analysis}, we compare the physical and evolutionary properties of galaxies with and without significant nebular contribution. Section \ref{sec:discussion} discusses the main results and their implications for future studies and in Sect. \ref{sec:sum_and_conc} we summarise the work. We adopt a cosmology with H$_{0}$ = 70 km s$^{-1}$ Mpc$^{-1}$, $\Omega_{M}$ = 0.3 and  $\Omega_{\Lambda}$ = 0.7, and a \cite{chabrier03} initial mass function (IMF) to estimate the star formation rate (SFR). Furthermore, whenever we mention EW measurements, we specifically mean their rest-frame values.

\section{Sample and methodology}
\label{sec:sample_methods}
    
To select our sample, we explored the DAWN JWST Archive (DJA)\footnote{\url{https://dawn-cph.github.io/dja/}}, a public repository of imaging and spectroscopic data from a variety of different observing programmes conducted with JWST. We focused on the spectroscopic data available in DJA (version 3), obtained using NIRSpec \citep{jakobsen22,ferruit22} across approximately 40 different observing programmes, and processed using MSAEXP \citep{brammer22,heintz24,degraff25}.

NIRSpec is a spectrograph onboard of the JWST and operates in the near-infrared regime ($\sim$0.6--5.3 $\mu$m), with three resolution modes (R$\sim$100, R$\sim$1000 and R$\sim$2700) obtained through different grating-filter combinations. Note that due to a physical gap in the detector, the obtained spectra might not be continuous and it may occur, depending on the redshift of the observed source, that relevant emission lines and spectral features are not observed.

The spectroscopic catalogue contains a total of 40691 spectra. However, considering that the same objects are observed using different instrument configurations and in different observing programmes, the actual number of unique sources is 21076. The catalogue has a quality flag related to the robustness of the derived redshift estimates. There are 30144 spectra classified with the highest quality flag (\texttt{grade=3}), corresponding to 12574 unique sources. This will be the sample considered in the following steps.

Since our objective is to carry out spectral fitting in the rest-frame optical regime, it is fundamental that the spectra enable the detection of both the continuum and the most prominent emission lines. These constraints lead to the consideration of only spectra obtained using the medium-resolution mode (R$\sim$1000), since the low-resolution mode does not provide sufficient resolving power to disentangle emission lines of interest that are close to each other, and the high-resolution mode leads to spectra with low S/N in the continuum. Additionally, to ensure a significant coverage of the rest-frame optical, we also impose that each source has observations for at least two consecutive grating-filter combinations. This leads to a selection of 3378 unique sources.

We aim to apply FADO to the selected sample. We give below a short overview of this spectral fitting tool \cite[for a detailed explanation, see][]{gomes17}. FADO aims to identify the linear combination of simple stellar populations (SSPs) that best replicates the stellar and nebular characteristics of a spectra. It fits the spectral continuum while masking emission lines, which separately fitted with Gaussians to measure flux and EWs. Simultaneously, FADO computes the nebular emission using an internal photoionisation routine which, when possible, also computes the electron temperature (T$_{e}$) and density (n$_{e}$), otherwise standard conditions are assumed: T$_{e}$ = 10$^{4}$ K and n$_{e}$=100 cm$^{-3}$. These processes are conducted on the fly while ensuring consistency between the stellar and nebular components.

Based on works that tested the capabilities of FADO to recover reliable physical properties \citep{cardoso19,pappalardo21}, we require that the spectra provide a significant wavelength coverage of the rest-frame optical, particularly of the Balmer break and relevant emission lines (H$\beta$, [OIII]$\lambda$5007, H$\alpha$, [NII]$\lambda$6584) and that they have a reasonable median signal-to-noise ratio (S/N $>$ 5). Following the application of these criteria, we performed a visual inspection of the selected spectra and obtained a sample of 143 unique sources. The main goal was to ensure that the position of the gap within the spectrum did not impact the detectability of the relevant emission lines, as their proximity could lead them to be inadequate for fitting.

Next, we joined the spectra of the same source obtained with different grating-filter combinations (G140M/F100LP, G235M/F170LP and G395M/F290LP), and thus different spectral coverage, into a single spectrum. This procedure requires a reliable flux calibration of the spectra obtained with each grating. This can be achieved using the overlapping region between the gratings \citep[e.g.][]{sanders25}.

In the overlapping region, we calculated the average flux and then identified the flux measurements that deviated more than 3$\sigma$ relative to this value. Then, we masked these points and recalculated the average flux of the overlapping region. By following this procedure for each grating, we obtain an estimate of the relative flux calibration between gratings. We used the G235M grating as a reference for the calibration, as it is the central grating and thus has an overlapping region with the other two gratings. The ratio of the average continuum flux in the overlapping region between G235M and each grating is used as the scale factor, resulting in a final ratio of one between the average continuum flux in the overlapping regions between gratings. The mean scale factor between gratings was 0.98 for G140M/G235M and 0.99 for G395M/G235M, comparable to literature results \citep[e.g.][]{sanders25}. Because the sources have different redshifts and different available gratings, the final spectra cover slightly different wavelength ranges. However, in all cases we ensure extensive coverage from the Balmer break to [NII]$\lambda$6584, apart from possible wavelength discontinuities due to the physical gap in the detector.

Before the spectral fitting procedure, we corrected the obtained spectra for Galactic extinction using the \cite{schlegel98} dust maps together with the correction factor determined by \cite{schlafly11}, and considering the \cite{cardelli89} extinction curve. Finally, we converted the spectra to the rest-frame.

For the FADO application, we used the same spectra basis and fitting specifications as in \cite{miranda25} (to ensure consistency between works), and we present a brief summary next. FADO was applied in the 3000--9000 \AA\ wavelength range, using SSPs from \cite{bruzual03} with a \cite{chabrier03} IMF and Padova 1994 evolutionary tracks \citep{alongi93,bressan93,fagotto94a,fagotto94b,girardi96}. The main set of SSPs is composed of 57 ages (t = [0.5 Myr, 13 Gyr]) and 3 metallicities (Z = 0.2, 0.4, 1 Z$_{\odot}$), including all SSPs in \cite{bruzual03} with ages between 0.5--15 Myr, to cover in detail the young SSPs which contribute most significantly to the overall nebular emission. From this main set of SSPs, we only selected those that are younger than the age of the Universe at the redshift of each individual galaxy. For dust attenuation corrections, we considered the \cite{calzetti00} extinction law.

To further trim the sample, we evaluated the quality of the fit based on the $\chi^{2}$ value\footnote{In this paper, when using $\chi^{2}$ we are referring to the reduced $\chi^{2}$.}, selecting fits with values lower than 5. Furthermore, with the flux measurements obtained by FADO, we selected galaxies classified as SF following the BPT diagram \citep{baldwin81}. Considering that the BPT diagram is known to misclassify low-metallicity active galactic nuclei (AGNs) (more common at high redshifts) as SF galaxies \citep[e.g.][]{groves06,kocevski23,harikane23,chisholm24,maiolino24}, we also removed galaxies that show evidence of a broad component in the H$\alpha$ region. Our aim with these criteria is to remove galaxies that might contain an AGN and obtain a reliable sample of SF galaxies to conduct the final analysis. In the end, we obtained a sample of 54 SF galaxies. As a consequence of our selection criteria, the sample size decreases significantly from the complete DJA spectroscopic catalogue to our final sample. However, this decrease is consistent with works that employ similar strict selection criteria to this database \citep[e.g.][]{liu26}.

Summarising, we followed the outlined procedure:

    \begin{itemize}
        \item Identification of unique sources in DJA.
        \item Selection of sources with robust redshift estimates (\texttt{grade=3}).
        \item Selection of galaxies with spectra obtained with NIRSpec mid-resolution mode (R$\sim$1000) for at least two consecutive grating-filter combinations.
        \item Selection of spectra with significant coverage of rest-frame optical wavelengths, including the Balmer break and BPT emission lines, and median S/N $>$ 5.
        \item Preliminary FADO fit and selection of well-fitted galaxies ($\chi^{2}$ $<$ 5), classified as SF following the BPT diagram and with no evidence of broad component in H$\alpha$ region.
    \end{itemize}

\begin{figure}[!ht]
        \centering
        \includegraphics[scale=0.36]{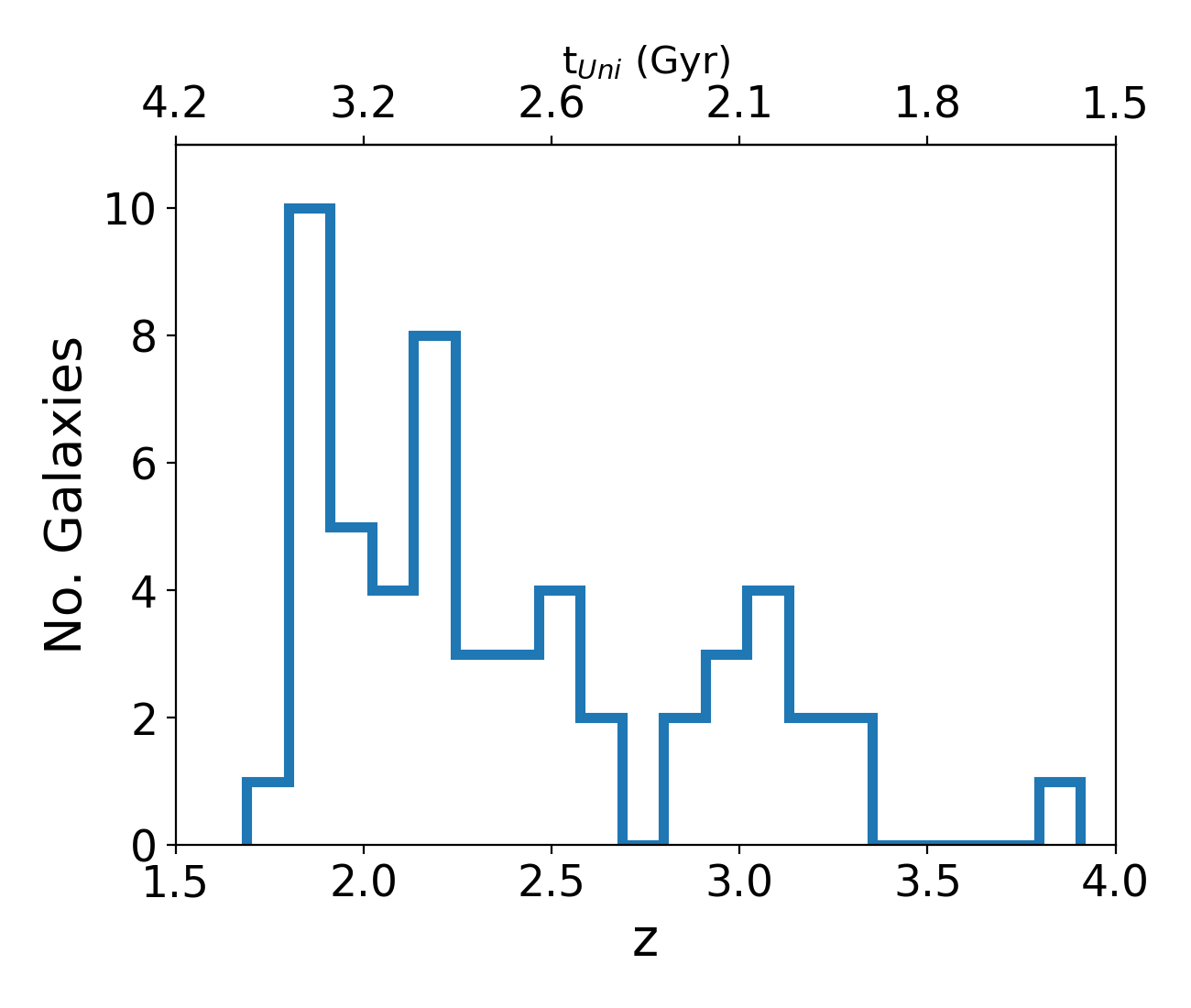}
        \caption{Redshift distribution of our sample, with the corresponding age of the Universe, in Gyr, indicated on the upper x-axis.}
        \label{fig:z_distribution}
    \end{figure}

In Fig. \ref{fig:z_distribution} we show the redshift distribution of the sample. The sample covers the 1.7 $<$ $z$ $<$ 3.9 range, with 29\% of the sample at $z$ $<$ 2, 53\% at 2 $<$ $z$ $<$ 3 and  18\% at z $>$ 3. This redshift range corresponds to an age of the Universe approximately between 1.5 and 3.5 Gyr. The low redshift limit of our sample arises from the necessity of capturing the Balmer break, and at lower redshifts this feature falls outside the wavelength coverage provided by NIRSpec. On the other hand, at higher redshift, the limiting factor is that it is increasingly difficult to obtain spectra that meet our quality criteria.
    
It is relevant to highlight that our sample is selected from a comprehensive repository which contains data from a variety of different observing programmes, each one designed with distinct scientific objectives. Furthermore, to obtain the sample, we focused solely on selecting galaxies with spectra that met the defined criteria, without attempting to select a cohesive group of galaxies. In this way, our sample is not representative of the overall galaxy population at the covered redshifts. Although this fact may limit the generalisation of the results, it does not impact our overall goals.

To obtain robust emission line measurements and physical and evolutionary properties estimates for the galaxies in the final sample, we perturbed the observed spectrum and executed the FADO run 100 times for each galaxy \citep[similar to the procedure followed in][]{miranda25}. Two separate sets of FADO runs were conducted, one in full-consistency (FC) mode and other in pure-stellar (PS) mode. While for the former both the stellar and nebular continuum are considered when fitting the spectrum, for the latter only the stellar continuum is taken into account. For the spectrum perturbation procedure, we assumed that the flux measurement error followed a Gaussian distribution centred at zero, with a standard deviation corresponding to the flux measurement error. The final estimate and associated uncertainty for each emission line measurement and the physical and evolutionary properties are derived from the mean and standard deviation calculated across all runs for each galaxy. We consider these estimates for the analysis presented in the following sections. Regarding the physical properties, we will focus specifically on the currently available and total ever formed stellar mass (M$_{curr}$ and M$_{ever}$, respectively), light- and mass-weighted stellar ages (t$_{L}$ and t$_{M}$, respectively) and light- and mass-weighted stellar metallicities (Z$_{L}$ and Z$_{M}$, respectively).

We note that FADO uses a non-standard EW definition, that on average estimates 0.1 dex higher EW values relative to the standard EW definition used in the literature \citep[for details see][]{miranda23}. Hence, the EW measurements presented in this work need to be scaled by this factor to be comparable with works that use the standard definition. For example, the threshold of EW(H$\alpha$) $\geq$ 500 \AA\ for significant impact on the derived properties of galaxies due to the self-consistent modelling of the stellar and nebular continuum defined in \cite{miranda25}, actually corresponds to EW(H$\alpha$) $\geq$ 375 \AA\ in the standard EW definition.

The spectra used in this work are derived from slit-based spectroscopy, and in some cases the slit size might not encompass the entire galaxy. We visually compared the slit dimension and position relative to the galaxies in our sample, and only in some cases the galaxies are nearly fully covered by the slit. Consequently, the spectra originate from a specific region, rather than from the complete galaxy. The consequences of these aperture effects have been studied \citep[e.g.][]{agostino23,dalmasso25} and efforts have been made to correct them \citep[e.g.][]{iglesiasparamo13,duartepuertas17}. However, addressing aperture effects is a complex subject that is beyond the scope of this work, and we do not correct our estimates for this effect.

\section{Fit quality and nebular continuum analysis}
\label{sec:FitQuality_NebCont}

\subsection{Assessment of the fit quality}

\begin{figure}[!ht]
        \centering
        \includegraphics[scale=0.36]{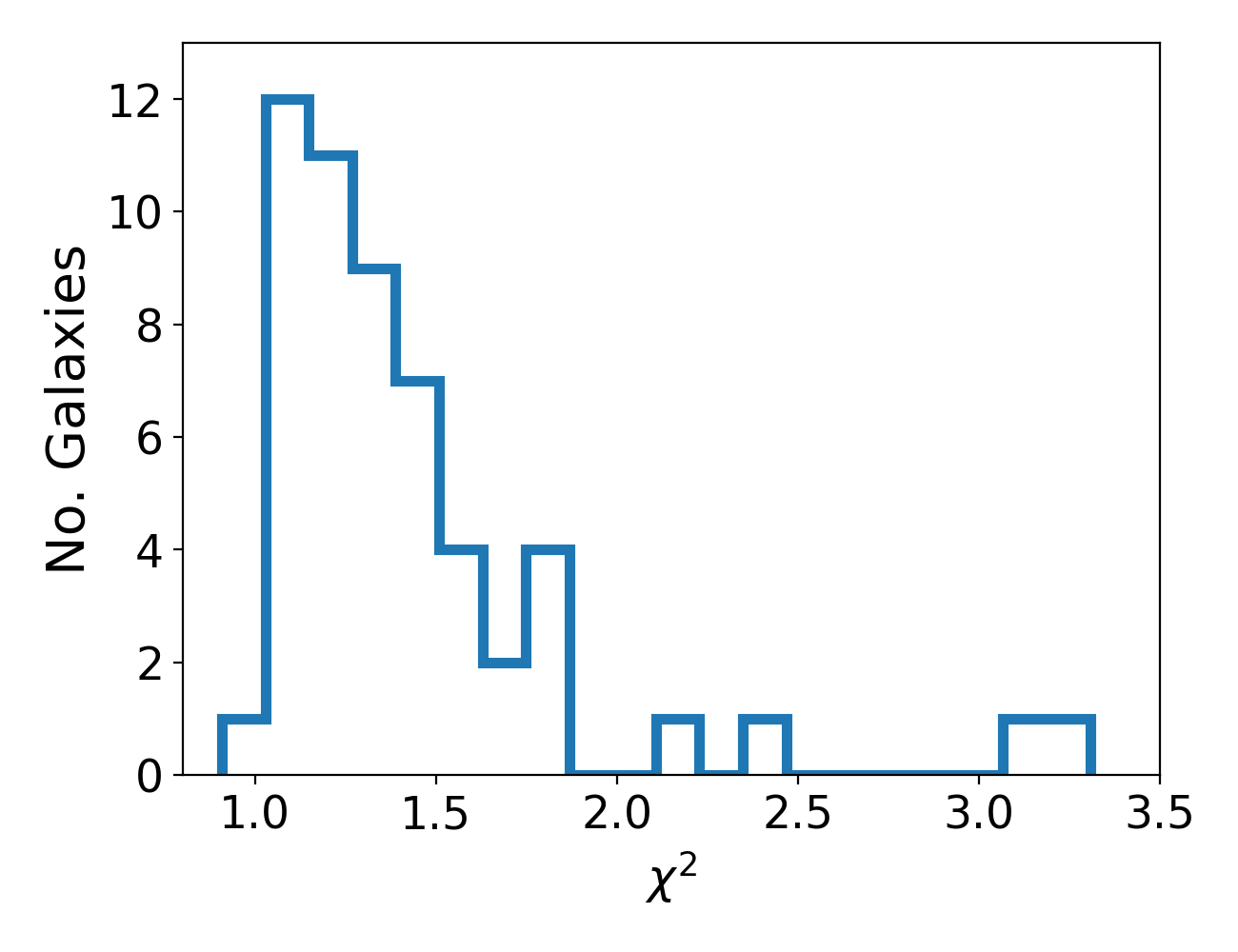}
        \caption{Distribution of the mean $\chi^{2}$ across 100 runs by FADO on the perturbed input spectrum of each galaxy in our sample.}
        \label{fig:chi2_distribution}
    \end{figure}

Before studying the derived results, it is important to evaluate the quality of the fits obtained by FADO. For each galaxy, we calculated the mean $\chi^{2}$ of the 100 runs obtained applying FADO to the perturbed input spectra. In Fig. \ref{fig:chi2_distribution} we show the distribution of the mean $\chi^{2}$ for our sample.

For all galaxies the obtained $\chi^{2}$ values are reasonable, as expected considering the selection criteria. The mean of the distribution is $\chi^{2}$=1.43 with the values being mostly concentrated in the 1 $<$ $\chi^{2}$ $<$ 2 interval. The galaxies with $\chi^{2}$ $>$ 2 were visually inspected to verify the adequacy of the fits, as they deviate from the other galaxies, and no issues were found. The mean standard deviation of the $\chi^{2}$ values for the 100 runs by FADO is 0.05. The small variation of the $\chi^{2}$ between each run on the perturbed spectra shows the stability of the solutions obtained by FADO for the galaxies in our sample.

The S/N of the spectra is an important factor contributing to this stability. The typical mean S/N in the continuum for our sample is $\sim$13 which is adequate to obtain reliable results using FADO \citep{pappalardo21}. Another relevant factor is restricting our sample to spectra obtained with the medium-resolution mode of NIRSpec, whose spectral resolution is sufficient to properly disentangle and model close spectral emission lines. This is fundamental for the adequate determination of H$\alpha$ and H$\beta$ luminosities and EWs, which are fundamental for the self-consistent estimation of the nebular continuum by FADO \citep{gomes17}. This result confirms that our selection criteria fulfilled the objective of selecting spectra with sufficiently high quality to carry out full spectral fitting with FADO.

\begin{figure}[!ht]
        \centering
        \includegraphics[scale=0.33]{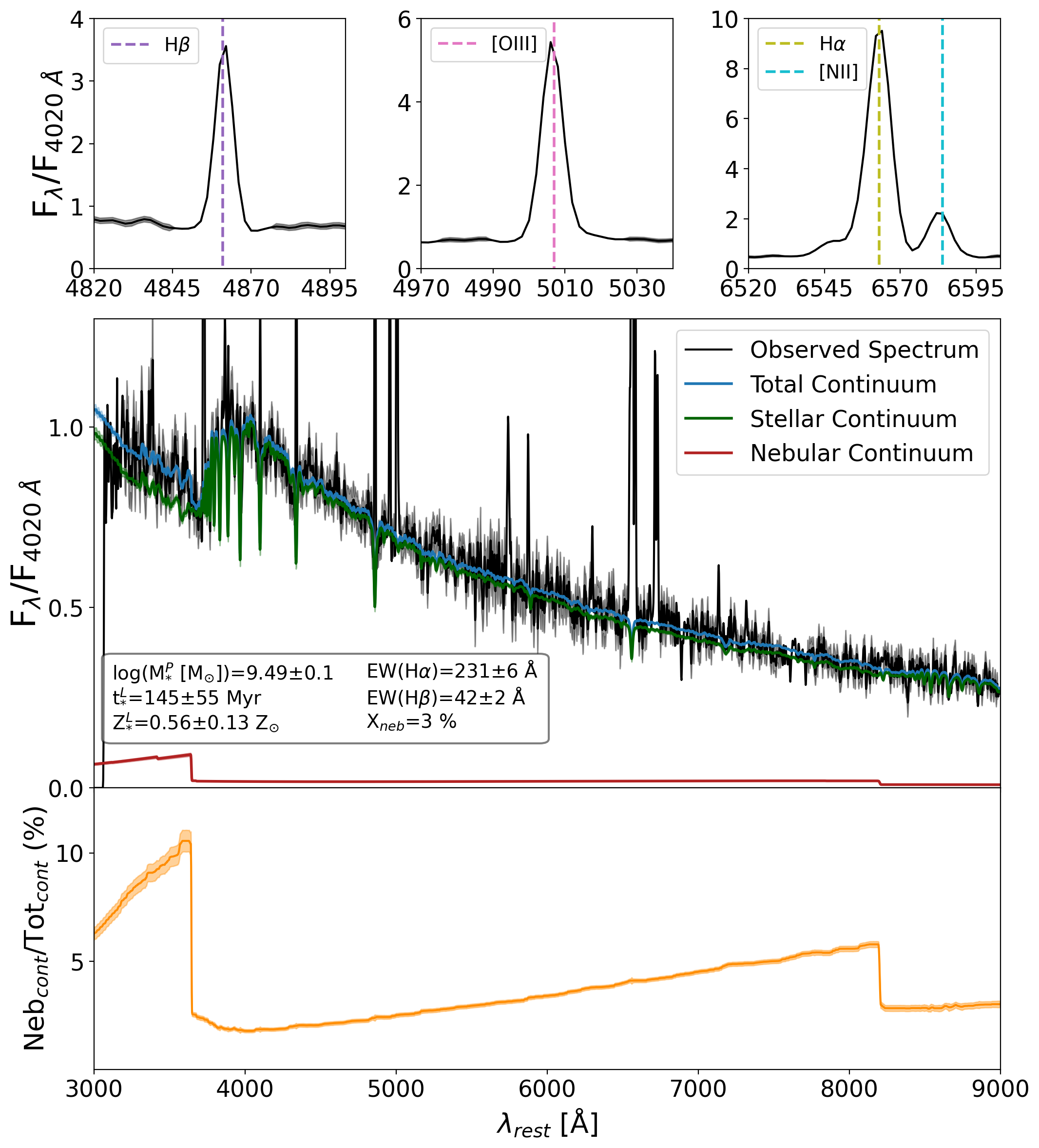}
        \caption{Model fitted by FADO to galaxy with S/N$\simeq$13, representative of typical mean value for our sample. Galaxy with DJA \texttt{uid} 13177 and $z$$\simeq$2.20, observed by the Blue Jay survey \citep{belli25}. \textit{Upper Panels:} Observed spectrum, plus zoom-in on BPT emission lines, (black line) and total, stellar and nebular continuum fitted by FADO (blue, green and red lines, respectively). The shaded regions represent the uncertainty in the observed spectrum and estimated models. The y-axis shows the normalised flux at $\lambda$=4020 \AA. The inset box shows FADO estimates for some properties and associated uncertainties: presently available stellar mass ($M_{*}^{P}$), light-weighted stellar age ($t_{*}^{L}$) and metallicity ($Z_{*}^{L}$), H$\alpha$ and H$\beta$ EW and nebular contribution (X$_{\text{neb}}$). \textit{Bottom Panel:} Ratio between the nebular and total continuum fitted by FADO as a function of wavelength, with the shaded region representing the estimated uncertainty.}
        \label{fig:Fit_Example}
    \end{figure}

In the upper panels of Fig. \ref{fig:Fit_Example}, we show a FADO fit to galaxy in our sample with S/N$\simeq$13 in the continuum, representative of the typical mean value for our sample, together with a zoom-in on the BPT emission lines. The filled lines represent the observed spectrum and fitted models, while the shaded regions represent the associated uncertainties. In the lower panel, we show the ratio between the nebular and total continuum as a function of wavelength.

We can see that the continuum is being accurately traced, thereby demonstrating a robust fit to the data. Furthermore, we can see that the estimated uncertainties associated with the physical properties are reasonable (the median relative error for the presently available stellar mass, light-weighted stellar age and metallicity are 29\%, 31\% and 16\%, respectively). This shows that the estimated values do not change significantly when perturbing the data following the associated errors. 

The interplay between the stellar and nebular continuum explains the wavelength dependence of the ratio between the nebular and total continuum. This ratio reaches its maximum at smaller wavelengths, blueward of the Balmer break, where the nebular continuum is strongest. After the sudden decrease redward of the Balmer break, the ratio gradually increases with wavelength, since the stellar continuum declines steeply and the nebular continuum remains approximately flat, before decreasing abruptly again at the Paschen break, due to the sudden decrease of the nebular continuum.

In Appendix \ref{appendix_FADO_Example}, we also show a FADO fit to a galaxy with S/N$\simeq$7 (Fig. \ref{fig:Fit_Example_Low}) and S/N$\simeq$30 (Fig. \ref{fig:Fit_Example_High}), representative of the galaxies in our sample with lowest and highest mean S/N in the continuum, respectively. Additionally, we show a FADO fit to galaxies with elevated X$_{\text{neb}}$ values, specifically X$_{\text{neb}}$=10\% (Fig. \ref{fig:Fit_Example_MidXneb}) and X$_{\text{neb}}$=21\% (Fig. \ref{fig:Fit_Example_HighXneb}).

\subsection{Characterisation of nebular contribution}
\label{subsec:Neb_cont}

In this work, we consider the definition of X$_{\text{neb}}$ introduced in \cite{miranda25}, defined as the median value of the ratio between the nebular and total continuum at each wavelength in the 3000--9000 \AA\ interval. This value provides a general measurement of the relative importance of the nebular continuum at optical wavelengths, although the contribution of the nebular to the total continuum changes with wavelength (see bottom panel of Fig. \ref{fig:Fit_Example}).

\begin{figure}[!ht]
        \centering
        \includegraphics[scale=0.36]{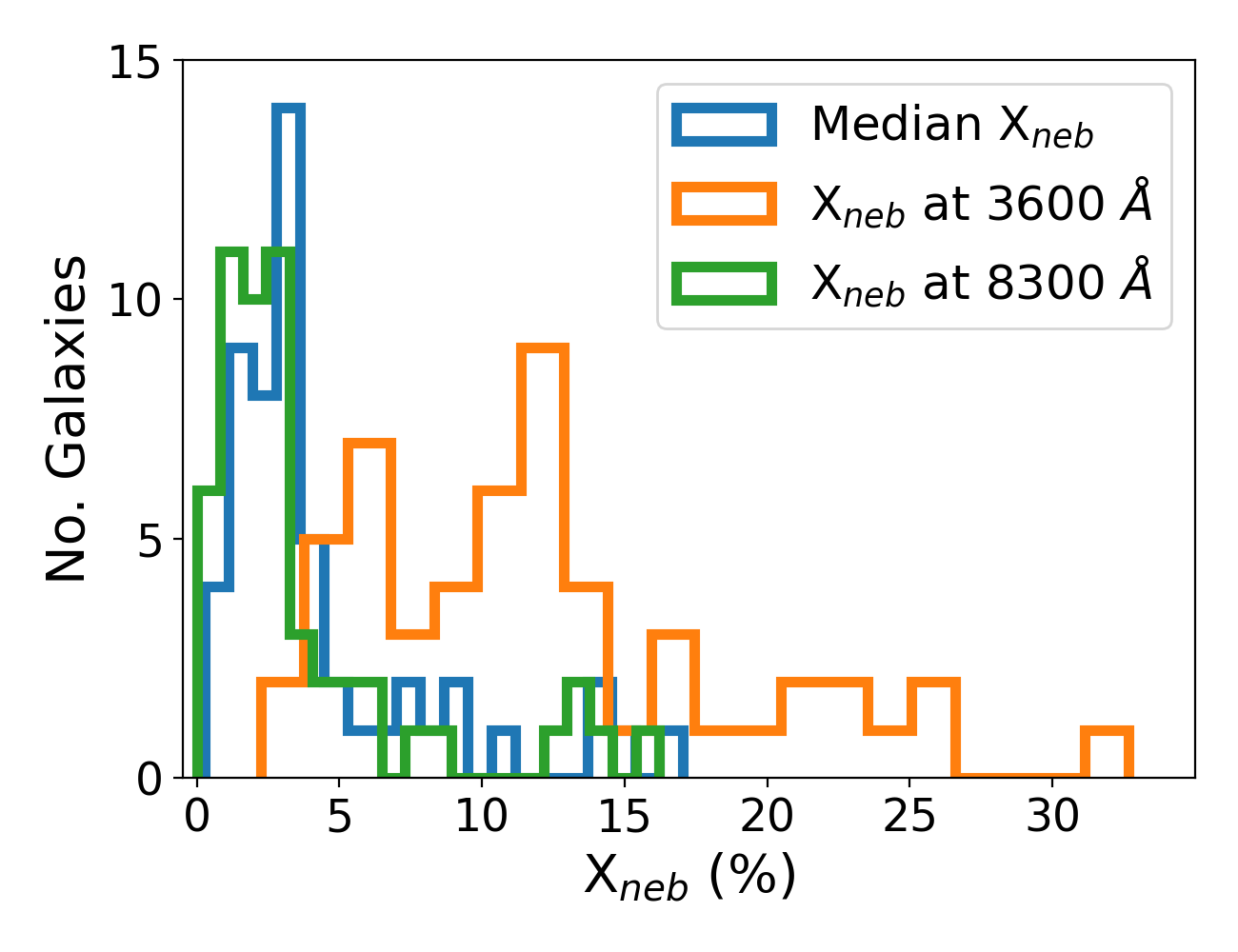}
        \caption{Distribution of the estimated median nebular contribution (blue) and at 3600 \AA\ (orange) and 8300 \AA\ (green) for the galaxies in our sample.}
        \label{fig:Xneb_distribution}
    \end{figure}

Figure \ref{fig:Xneb_distribution} shows the X$_{\text{neb}}$ distribution and also the nebular contribution at 3600 \AA\ and 8300 \AA, blueward of the Balmer discontinuity and redward of the Paschen discontinuity, respectively. We select these two wavelengths as they represent the regions where the nebular contribution in the optical spectral range is approximately at its maximum and minimum, respectively, as it can be seen in the lower panel of Fig. \ref{fig:Fit_Example}.

The mean nebular contribution of our sample is X$_{\text{neb}}$$\simeq$4\%, and the distribution is strongly skewed towards lower values, with 76\% of the sample having X$_{\text{neb}}$ $<$ 5\%. Furthermore, we see that the distribution for the nebular contribution at 8300 \AA\ is comparable to that of X$_{\text{neb}}$, whereas the nebular contribution at 3600 \AA\ is shifted towards comparatively higher values and with a larger dispersion.

Comparing the X$_{\text{neb}}$ distribution with the results from \cite{miranda25}, we find that the general trend is consistent between the works. The main difference is the amount of galaxies with high X$_{\text{neb}}$ values. While in \cite{miranda25} it is reported that 14\% of the sample has X$_{\text{neb}}$ $>$ 15\%, for our sample only 2\% (one galaxy) meet the same condition. Furthermore, considering that the star-forming activity of the Universe was higher in the past, it would be expected that the fraction of galaxies with high X$_{\text{neb}}$ values was also higher \citep{madau14,faisst16,boyett24,miranda25}, the opposite of what our results suggest.

Galaxies with such high X$_{\text{neb}}$ values (EW(H$\alpha$) $>$ 1000 \AA) are extremely rare, both at low and high redshifts \citep[e.g.][]{amorin15,withers23,llerena24,boyett24}. However, the sample used in \cite{miranda25} was selected considering the EW(H$\alpha$) and specifically designed to contain such galaxies, not being representative of the complete SDSS sample from which is drawn. On the other hand, for the sample used in this work, we merely apply conditions related to spectral quality, line detectability and consider no EW(H$\alpha$) selection, leading to a sample not representative of galaxy population at the covered redshifts (see Sect.\ref{sec:sample_methods} for details). This justifies the difference between the two samples in the amount of galaxies with high X$_{\text{neb}}$ values and the contrary evolution from what would be expected.

In \cite{miranda25}, the EW of H$\alpha$ and H$\beta$ were identified as reliable tracers of X$_{\text{neb}}$, exhibiting a tight and positive linear correlation with this quantity in log-log space ($\rho$ $>$ 0.99). The specific SFR (sSFR) was also identified as a tracer of X$_{\text{neb}}$, albeit with a substantially larger intrinsic scatter in the corresponding relation ($\rho$$=$0.84). Even though our sample is not very large, it is still interesting to assess how it compares with these previously established trends. 

We estimate the SFR from the H$\alpha$ flux following the same procedure as in \cite{miranda25}, to ensure comparability of results. We correct the H$\alpha$ flux for extinction through the Balmer decrement (H$\alpha$/H$\beta$) and then calculate the H$\alpha$ luminosity. Finally, we estimate the SFR using the \cite{kennicutt98} conversion factor calibrated to a \cite{chabrier03} IMF, $\eta$(H$\alpha$)=10$^{41.31}$ erg s$^{-1}$ M$_{\odot}^{-1}$ yr. We use this SFR estimate, together with the stellar mass derived by FADO to estimate the sSFR.

\begin{figure*}[!ht]
        \centering
        \includegraphics[scale=0.4]{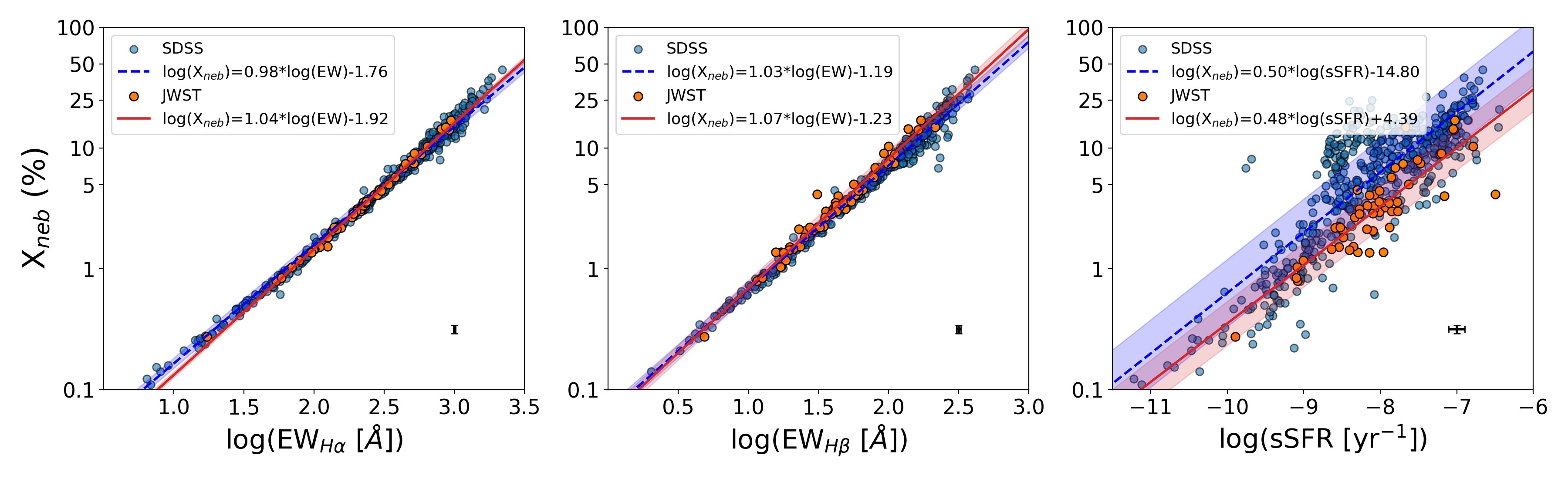}
        \caption{Relation between X$_{\text{neb}}$ and the EW of H$\alpha$ (left panel), EW of H$\beta$ (middle panel) and sSFR (right panel). The orange points are the JWST sample considered in this work and the red line is the corresponding linear fit. The blue points are the SDSS sample from \cite{miranda25}, and the blue dashed line shows the corresponding linear fit. The shaded regions represent the 1$\sigma$ scatter for each linear fit. The median error of the JWST data is presented in the bottom right corner. The axes are in logarithmic scale.}
        \label{fig:Xneb_vs_tracers}
    \end{figure*}

In Fig. \ref{fig:Xneb_vs_tracers}, we show the relation between X$_{\text{neb}}$ and the EW of H$\alpha$ and H$\beta$ and sSFR for our JWST data, plus the linear fit to the data. Additionally, we plot the SDSS data and derived linear fit from \cite{miranda25}. The shaded regions represent the 1$\sigma$ scatter for each linear fit.

The JWST and SDSS data show a clear agreement for the relation between X$_{\text{neb}}$ and the EW of H$\alpha$ and H$\beta$. In fact, we see that the 1$\sigma$ scatter of both linear fits overlap significantly within the common range of the samples. We highlight that although H$\beta$ is intrinsically weaker than H$\alpha$, it is a valuable tracer of X$_{\text{neb}}$ when H$\alpha$ is redshifted out of an instrument’s spectral coverage (for NIRSpec at z $\gtrsim$  7).

The consistency of results between these two studies aligns with our expectations, considering the role of the EW of H$\alpha$ and H$\beta$ in the determination of the nebular continuum by FADO \citep{gomes17}. Despite the significant difference in the ages of the Universe spanned by the JWST and SDSS samples, and also the evolving conditions of the interstellar medium (ISM) between these epochs, it is interesting that the established relation remains consistent. When possible, FADO estimates the ISM properties (namely, T$_{e}$ and n$_{e}$) and considers them when estimating the nebular continuum. For our sample, FADO was unable to estimate these properties for 19 galaxies and reverted to the standard conditions. However, for the sample for which they were determined, it was obtained a mean T$_{e}$ $\simeq$ 10,200 K and n$_{e}$ $\simeq$ 150 cm$^{-3}$, comparable to the standard conditions. Thus, we are probing galaxies with ISM properties similar to the standard ones and some care must be taken when applying these relations to galaxies with more extreme conditions \citep[e.g.][]{curti25,zavala25,usui25}. Specifically, this discrepancy has been observed in \cite{reumert26}.

The EW values of the SDSS sample span approximately two orders of magnitude, whereas the JWST sample extends over only one order of magnitude, predominantly at the higher end of the distribution. This difference is also reflected in the respective coverage of X$_{\text{neb}}$: SDSS exhibits X$_{\text{neb}}$ values ranging from below 1\% to nearly 50\%, whereas the JWST sample only spans from a few percent up to approximately 25\%. This difference can be justified considering the details of how the SDSS and JWST samples were selected. Specifically, the requirement for continuum detection restricts the EWs of the JWST sample to lower values, being the maximum value EW(H$\alpha$) = 941 \AA, while for the SDSS sample is EW(H$\alpha$) = 2198 \AA.

On the other hand, there is a clear difference in the relation between X$_{\text{neb}}$ and sSFR for JWST and SDSS. The linear fits show comparable slopes, but a clear difference in the normalisation value. In fact, although the JWST data is in a region of the parameter space that is also covered by SDSS data, almost all points are below the linear fit to the SDSS data. This is further evidenced by the small overlap between the 1$\sigma$ scatter for the linear fits. This means that for the same value of X$_{\text{neb}}$, the sSFR tends to be higher for the JWST galaxies than for the SDSS ones. This could be an effect of the higher SF activity of the Universe at higher redshift \citep[e.g.][]{madau14}. However, considering the already described sample limitations, we cannot confirm that this is driven by this evolution or merely a selection effect. Additionally, we identify a subgroup of SDSS data (galaxies in the -9 $<$ log(sSFR) $<$ -8.5 interval and above the linear fit) that lead to a higher normalisation value for the linear relation derived in \cite{miranda25}.

Recent observations with JWST have enabled robust measurements of the Balmer discontinuity of high-redshift galaxies \citep[e.g.][]{vikaeus24,roberts-borsani24,katz25} and several results from both simulations and observations suggest that pronounced Balmer discontinuities become increasingly common toward higher redshifts \citep[e.g.][]{wilkins24,trussler25}. In particular, some 'nebular-dominated galaxies' have been identified with spectra that show a strong Balmer jump \citep[e.g.][]{cameron24,katz25}. In this way, this spectral feature could be an additional tracer of X$_{\text{neb}}$ at higher redshifts.

\begin{figure}[!ht]
        \centering
        \includegraphics[scale=0.35]{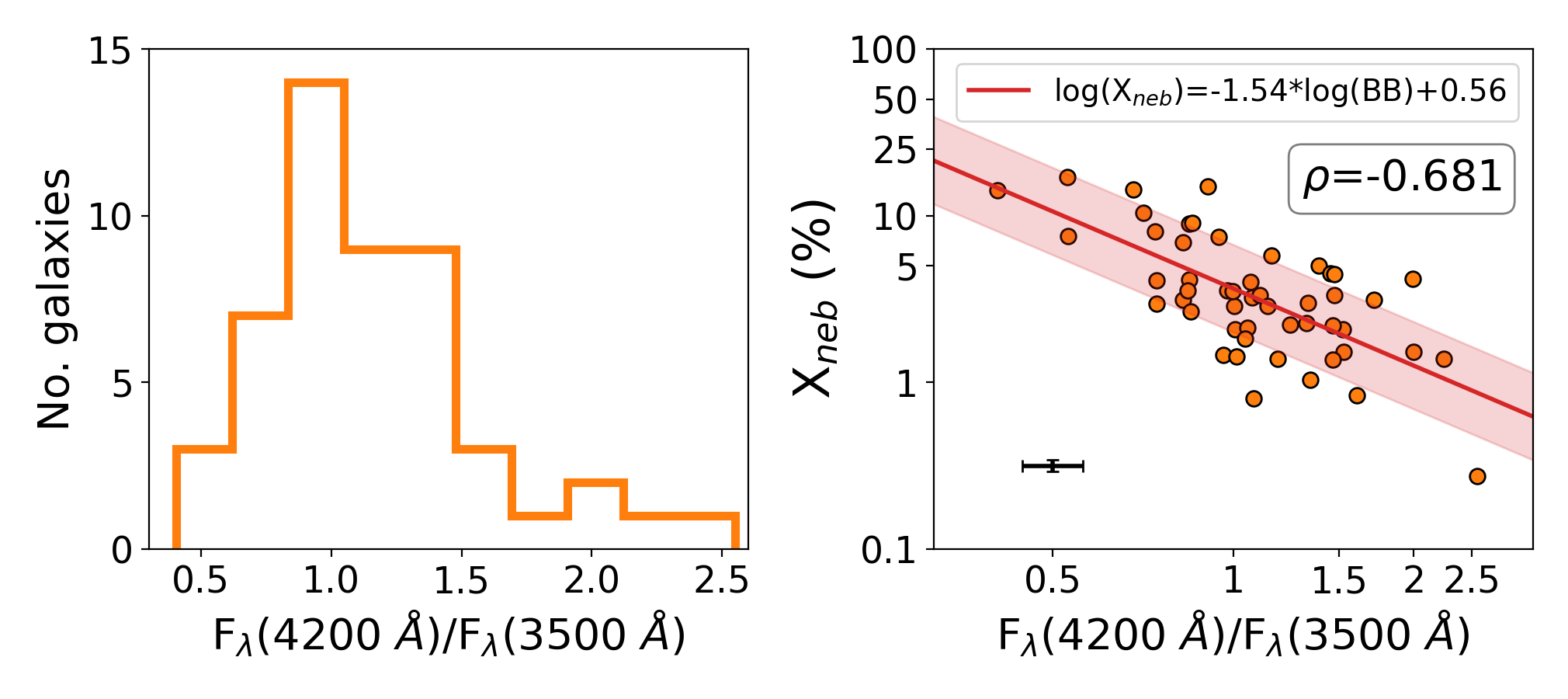}
        \caption{Distribution of the estimated Balmer break strength (left panel) and its relation with X$_{\text{neb}}$ (right panel). In the right panel, the red line shows the linear fit to the data and the shaded region the corresponding 1$\sigma$ scatter. We also show the best-fitting equation, the Pearson correlation coefficient and the median error of the data. The axes are in logarithmic scale.}
        \label{fig:Xneb_vs_BB}
    \end{figure}

We test this hypothesis using the the Balmer break strength as defined by \cite{binggeli19} B$_{4200/3500}$ = F$_{\lambda}$(4200 \AA) / F$_{\lambda}$(3500 \AA). We note that the Balmer break strength is usually measured using flux in units of F$_{\nu}$ rather than F$_{\lambda}$ \citep[e.g.][]{binggeli19,vikaeus24}. However, the estimates relate by B$_{\nu}$  $\sim$ 1.44 B$_{\lambda}$. For 4 galaxies it was not possible to obtain an estimate as the necessary wavelength region coincided with the spectral gap. In Fig. \ref{fig:Xneb_vs_BB} we show the distribution of the estimated Balmer break strength and its relation with X$_{\text{neb}}$.

For 40\% of the galaxies in our sample, the Balmer break strength is lower than 1 which means that the flux density is greater blueward of the break than redward, indicating spectra with relatively blue spectral slopes. This might indicate that for these galaxies the nebular continuum is sufficiently strong so that a Balmer jump is noticeable in the spectrum. In fact, we see that these galaxies are associated with larger X$_{\text{neb}}$ values. Regarding the relation between X$_{\text{neb}}$ and Balmer break strength, we find a moderate negative correlation ($\rho$ = -0.68). This weaker correlation with X$_{\text{neb}}$, compared to that seen for the H$\alpha$ and H$\beta$ EWs, may partly arise from dust attenuation effects, which impacts the Balmer break strength, due to the broader wavelength interval used in the estimation, but does not affect the EW estimates. Thus, the Balmer break strength can be used as a tracer of X$_{\text{neb}}$, even though the relation shows some scatter.

\begin{figure*}[!ht]
        \centering
        \includegraphics[scale=0.4]{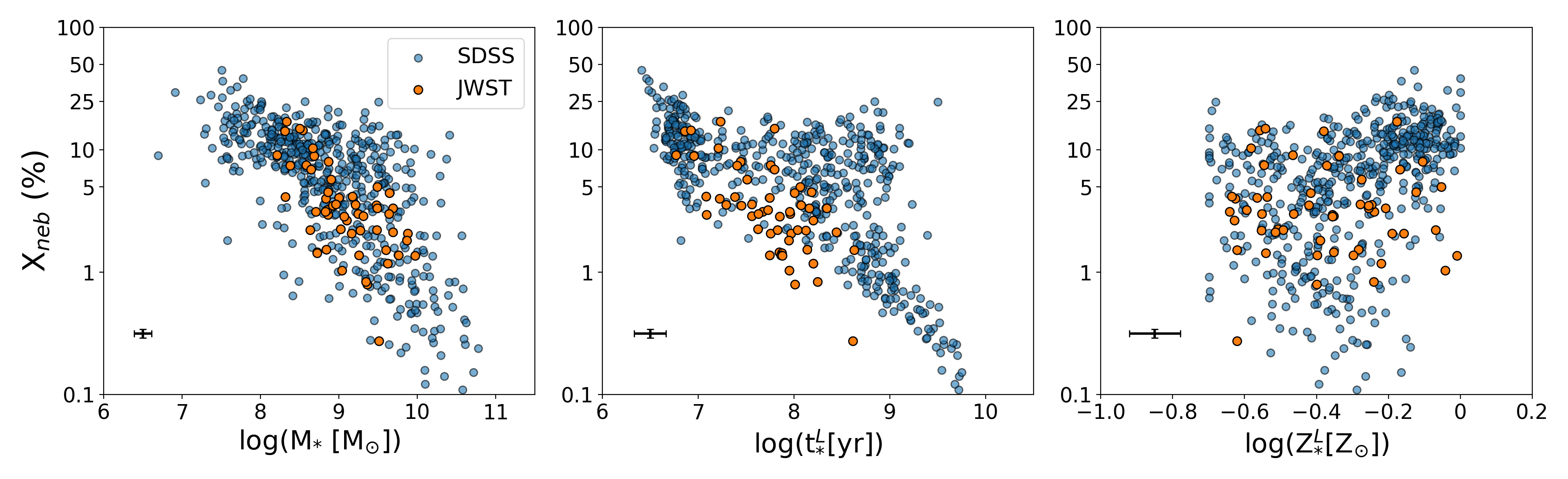}
        \caption{Relation between X$_{\text{neb}}$ and currently available stellar mass (left panel), light-weighted stellar age (middle panel) and light-weighted stellar metallicity (right panel). The orange points are the JWST sample considered in this work and the blue points are the SDSS sample from \cite{miranda25}. The median error of the JWST data is presented in the bottom left corner. The axes are in logarithmic scale.}
        \label{fig:Xneb_vs_galprop}
    \end{figure*}

In Fig. \ref{fig:Xneb_vs_galprop}, we analyse how X$_{\text{neb}}$ relates to relevant physical properties of galaxies, specifically the stellar mass, age and metallicity, and compare the results with those presented in \cite{miranda25}.

As reported in \cite{miranda25} for the SDSS data, we also observe an anti-correlation between X$_{\text{neb}}$ and stellar mass and age for the JWST data, with higher X$_{\text{neb}}$ values associated with low-mass and young galaxies. For the stellar mass, the parameter space covered by the JWST data is comparable to that of the SDSS data, although it does not extend to such low- (log(M$_{*}$ [M$_{\odot}$]) $<$ 8) or high-mass (log(M$_{*}$ [M$_{\odot}$]) $>$ 10) galaxies. Once again, this difference between the SDSS and JWST sample arises from the way in which the samples were obtained. For the stellar age, there is a similarity in the parameter space coverage between JWST and SDSS data, even though for the former the established relation is steeper. The age of the Universe for the JWST galaxies is younger than for the SDSS galaxies, so galaxies with comparable X$_{\text{neb}}$ values between SDSS and JWST will tend to have younger ages in JWST. The stellar metallicity coverage is similar for the JWST and SDSS samples, and no correlation with X$_{\text{neb}}$ is observed (similarly to \citealt{miranda25}).

\subsection{Nebular component impact threshold}

In \cite{miranda25}, a threshold of X$_{\text{neb}}$$\simeq$8\% (corresponding to EW(H$\alpha$)=500 \AA) was derived, above which the nebular continuum contribution becomes significant, and neglecting this contribution significantly biases the determination of the physical and evolutionary properties of galaxies. We analyse if this threshold still holds for our JWST data, by comparing the physical properties estimated by FADO when using its FC and PS modes.

We divided the sample into three bins: EW(H$\alpha$) $<$ 100 \AA, 100 $\leq$ EW(H$\alpha$) $<$ 500 \AA\ and 500 $\leq$ EW(H$\alpha$) $<$ 1000 \AA, obtaining 9, 38 and 8 galaxies for each bin, respectively. In \cite{miranda25} it was also considered a fourth bin, EW(H$\alpha$) $\geq$ 1000 \AA, but in our sample there are no galaxies in this interval. Despite the low number of galaxies in two of the three bins, the results can still be indicative of the impact of accounting for the nebular contribution.

\begin{figure}[!ht]
        \centering
        \includegraphics[scale=0.45]{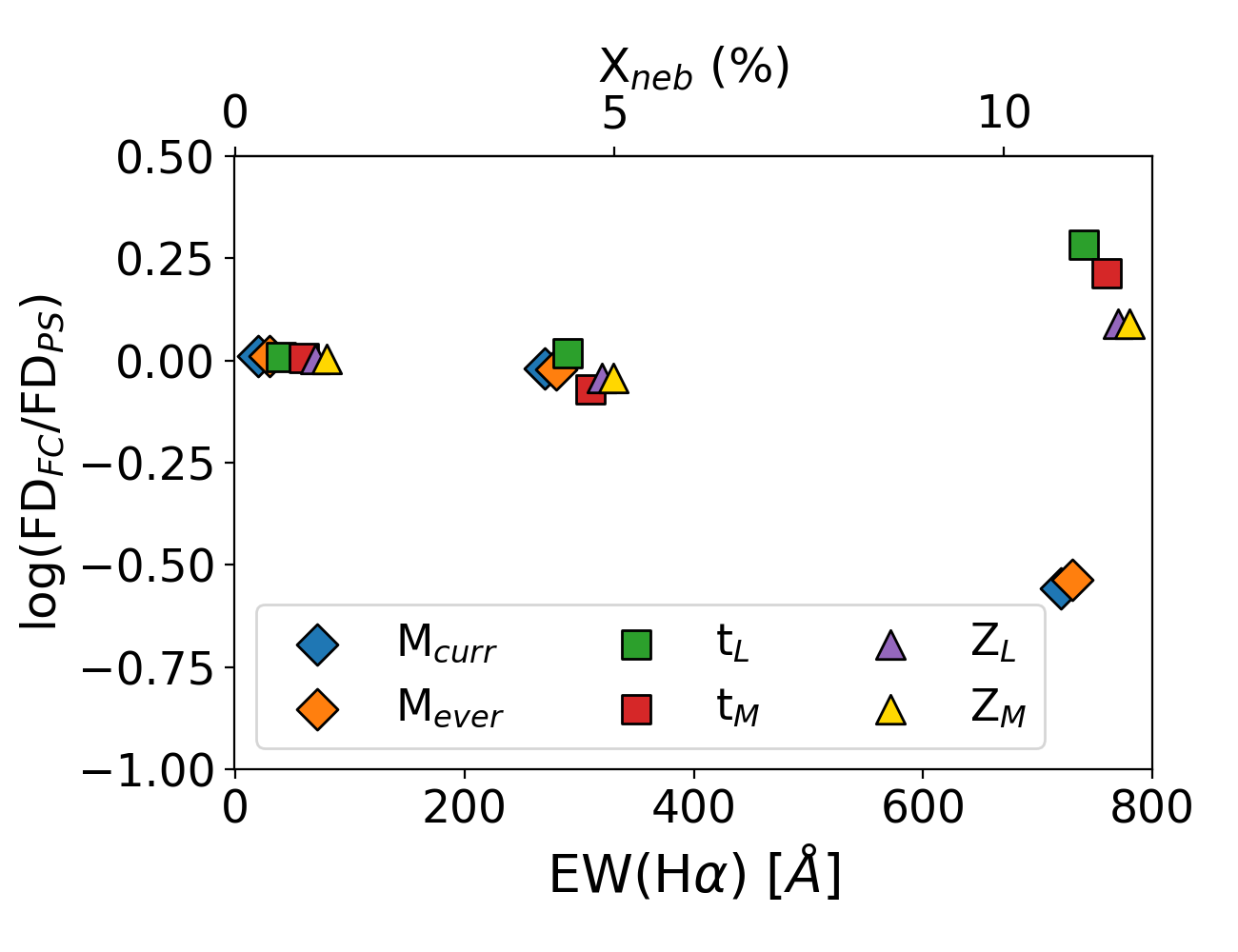}
        \caption{Logarithm of the median differences between FADO estimates in full-consistency mode (FD$_{FC}$) and in pure-stellar mode (FD$_{PS}$) for the stellar mass (diamonds), age (squares) and metallicity (triangles) for the previously considered EW(H$\alpha$) bins. Points in each bin are slightly shifted for clearer view.}
        \label{fig:FD_FC_vs_PS}
    \end{figure}

Figure \ref{fig:FD_FC_vs_PS} shows the logarithm of the median differences between the estimates obtained by FADO in FC and PS mode for the stellar mass, age and metallicity for each of the defined bins. The full distribution of the differences between the two modes for the considered physical properties are shown in Appendix \ref{appendix_FADO_FCvsPS}. In this comparison, we adopt a conservative uncertainty of 0.2 dex for the physical properties derived with FADO \citep{gomes17,cardoso19,pappalardo21}. 

In the first two bins all physical properties are within the 0.2 dex uncertainty interval, meaning that the estimates are not significantly different between the two FADO modes. For the third bin, 500 $\leq$ EW(H$\alpha$) $<$ 1000 \AA, the stellar mass estimates between the two modes become significantly different, with the PS mode obtaining values $\sim$0.5 dex larger than the FC mode. The differences between the two FADO modes for the stellar age estimates also become non-negligible, with the FC mode obtaining values $\sim$0.25 dex larger than the PS mode. On the other hand, the differences in the stellar metallicity estimates remain within the uncertainty level. 

The differences in estimated physical properties between the two modes arise from including the nebular continuum \citep[e.g.][]{izotov11,gomes17,breda22,cardoso22,miranda23,miranda25}. When both stellar and nebular continuum emission are considered, the stellar continuum is lower (since the total continuum is split between the stellar and nebular contribution) and the fit does not need to rely solely on older stellar populations to reproduce a flat continuum (since the nebular continuum is approximately flat, in terms of $f_{\lambda}$, at optical wavelengths). Therefore, for galaxies with significant nebular continuum emission, modelling both continua leads to the selection of different stellar populations relative to when the nebular continuum is neglected, leading to changes in the estimated physical properties.

The impact of the extra inclusion of the nebular continuum can also be directly visible when comparing the continua fitted by the FC and PS modes (see Appendix \ref{appendix_FADO_FCvsPS}). For a galaxy with low nebular contribution (X$_{\text{neb}}$ $<$ 1\%) we see that there is almost no difference between the FC and PS continua. However, for a galaxy with high nebular contribution (X$_{\text{neb}}$ $\simeq$ 8\%), the PS continuum fails to reproduce the spectral features characteristic of the nebular continuum, specifically the Balmer and Paschen discontinuities, whereas the FC mode reliably traces those features. This is coherent with the results reported by \cite{pappalardo21}, which were obtained by fitting mock spectra using FADO and other pure-stellar codes and comparing the results (see Fig. 14 of their work).

These results align well with the findings from \cite{miranda25}, being consistent with the same threshold limit for significant impact on the derived physical properties due to the modelling of both stellar and nebular continuum: EW(H$\alpha$) $\geq$ 500 \AA. Moreover, the magnitude of the difference between the estimates of the two modes is similar between the works. Considering the defined threshold, from the 54 galaxies in our sample, 8 exhibit EW(H$\alpha$)$\geq$500 \AA\ and are classified as having a significant nebular contribution.

\section{The properties of galaxies with significant nebular contribution}
\label{sec:HEW_galaxies_analysis}

In this section, we analyse and compare the physical and evolutionary properties between galaxies with and without a significant nebular contribution. Besides the JWST sample, we also include in the comparison the SDSS data from \cite{miranda25}, which contains 242 and 258 galaxies with and without significant nebular contribution, respectively. In what follows, galaxies with significant nebular contribution (EW(H$\alpha$) $\geq$ 500 \AA) are labelled as high EW (HEW), while those with lower values are referred to as low EW (LEW).

\begin{figure*}[!ht]
        \centering
        \includegraphics[scale=0.51]{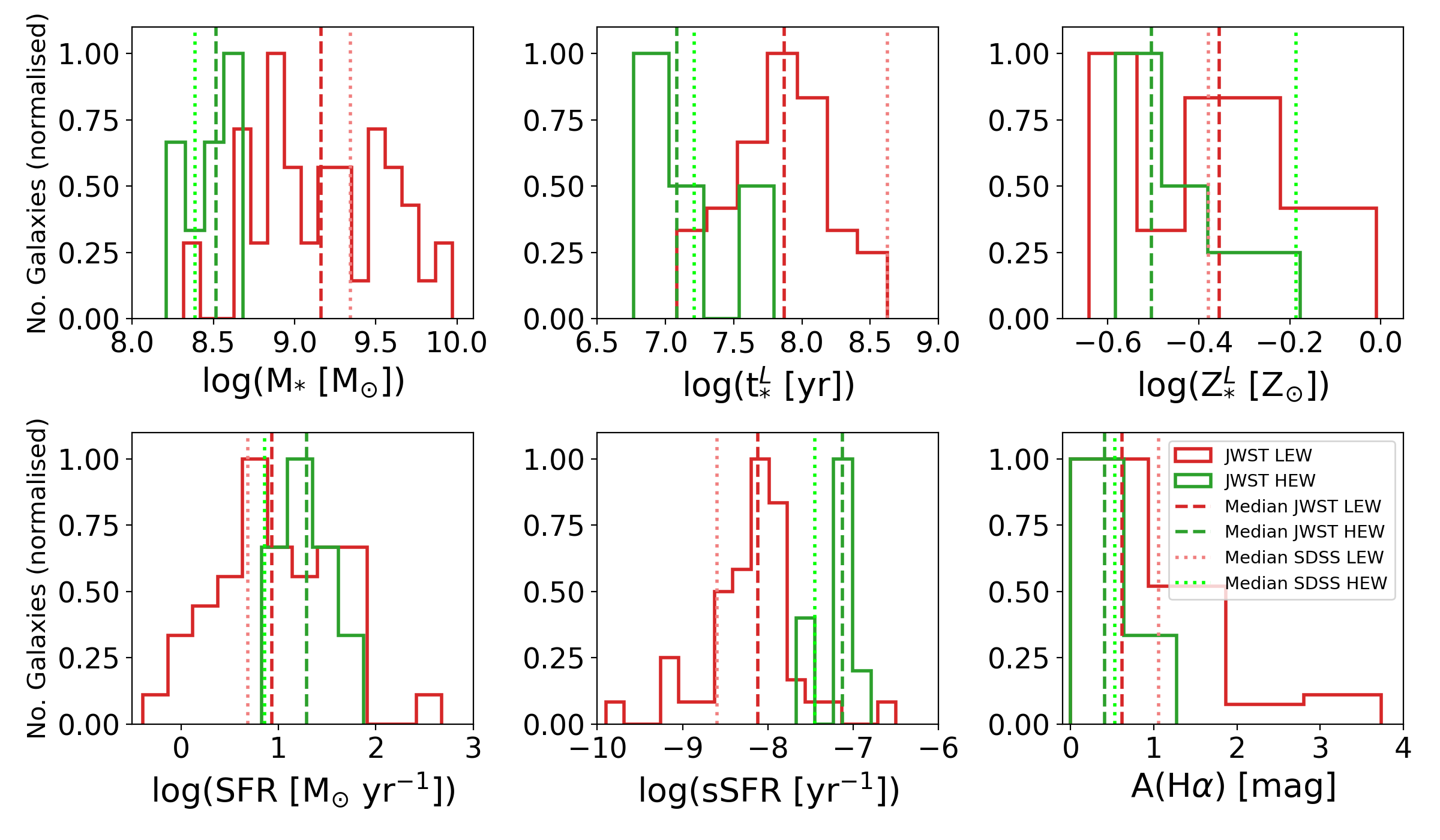}
        \caption{Distribution of the physical properties for the LEW and HEW subsamples (red and green, respectively) in the JWST data. The distributions are normalised so that the maximum value is equal to one. The vertical dashed lines indicate the median values of these distributions, while the dotted lines represent the median values for the corresponding subsamples from the SDSS data used in \cite{miranda25}. The upper row displays the currently available stellar mass and light-weighted age and metallicity from left to right, while the lower row shows the star formation rate, specific star formation rate and dust extinction from left to right.}
        \label{fig:Prop_Abv_Bel_Thresh_Gal}
    \end{figure*}

In Fig. \ref{fig:Prop_Abv_Bel_Thresh_Gal}, we present the distributions and median values of various physical properties for the HEW and LEW galaxies from the JWST sample. We also show the median value for the corresponding subsamples from the SDSS data used in \cite{miranda25}.

Both the stellar mass and age distribution for the HEW galaxies are concentrated at the lower end of the LEW galaxies distribution. The JWST HEW galaxies exhibit a maximum value of 10$^{8.7}$ M$_{\odot}$ and 10$^{7.8}$ yr for the stellar mass and age, respectively. In contrast, the distribution for LEW galaxies spans a considerably broader range, extending up to about 10$^{10}$ M$_{\odot}$ and 10$^{8.6}$ yr, for the stellar mass and age, respectively. This disparity is further highlighted by the difference between the median values of the two subsamples: $\approx$0.6 dex and $\approx$0.8 dex, for the stellar mass and age, respectively. 

When comparing the SDSS and JWST samples, we find that the median stellar mass for SDSS is somewhat lower for HEW galaxies (0.13 dex), and higher for LEW galaxies (0.18 dex). For the stellar age, the median values are larger for SDSS for both subsamples, 0.13 dex for HEW galaxies and grows to 0.75 dex for LEW galaxies. For the stellar mass, these results can be explained in light of the selection criteria of the two samples, whereas for the stellar age, the main driver is the younger age of the Universe for the JWST sample (see Sect. \ref{subsec:Neb_cont}).

For the stellar metallicity, HEW galaxies have on average 0.15 dex lower values  relative to LEW ones, even if the estimated values for LEW galaxies show larger dispersion, covering the full range of possible metallicity values (considering the spectral basis used). This is particularly noteworthy for the lowest metallicity values, which are compatible with the distribution for the HEW galaxies.

On the contrary, in the SDSS sample HEW galaxies have $\approx$0.2 dex higher stellar metallicities relative to LEW galaxies, even if the estimated values also cover the entire parameter range. Moreover, we see that the median value for the LEW galaxies is comparable between JWST and SDSS, but for HEW galaxies, the median value for JWST is $\approx$0.8 dex lower relative to SDSS. This suggests that this difference between JWST and SDSS is driven by the change in the estimated metallicity values for the sample of HEW galaxies.

For the SFR, there is an overlap between the distributions for the LEW and HEW galaxies, with the former covering approximately the 1 $<$ log(SFR) $<$ 2 range, while the latter spans a larger range, -0.5 $<$ log(SFR) $<$ 3. Nevertheless, the median value for the HEW galaxies is $\approx$0.4 dex higher relative to the LEW galaxies.

The distribution of the sSFR for the HEW galaxies locates at the high end of the distribution for the LEW galaxies. While the minimum sSFR value for the HEW galaxies is $\approx$10$^{-7.67}$ yr$^{-1}$, the distribution of the LEW galaxies spans to almost 10$^{-10}$ yr$^{-1}$. This tendency is further highlighted by the median values, which have approximately 1 dex difference between the two subsamples. 

The SFR and sSFR median values for the JWST sample are higher relative to the SDSS sample, for both LEW and HEW galaxies. This can be explained by the redshift range of the two samples and the well-known evolution of the SFR density of the Universe \citep[e.g.][]{madau14}. Similarly to the JWST sample, for SDSS the SFR and sSFR median values are higher for the HEW galaxies. In the case of the SFR, the difference between the JWST and SDSS sample is above 0.25 dex, whereas for the sSFR the difference is slightly higher, being above 0.32 dex. 

The dust extinction affecting H$\alpha$, A(H$\alpha$), for the LEW galaxies covers almost the entire parameter space (between 0 to 4 mag), with the distribution peaking at A(H$\alpha$) $<$ 1 mag. On the other hand, for the HEW galaxies, the distribution is almost completely in the A(H$\alpha$) $<$ 1 mag region. The difference of $\approx$0.2 mag between the median values also underscores the similarity between the distribution of the two subsamples.

When comparing A(H$\alpha$) between the JWST and SDSS samples, we see that for the HEW galaxies there is only a 0.13 dex difference between the median values, whereas for the LEW galaxies, the median difference is 0.44 dex. In both cases, the SDSS sample has higher median values.

\begin{figure*}[!ht]
        \centering
        \includegraphics[scale=0.55]{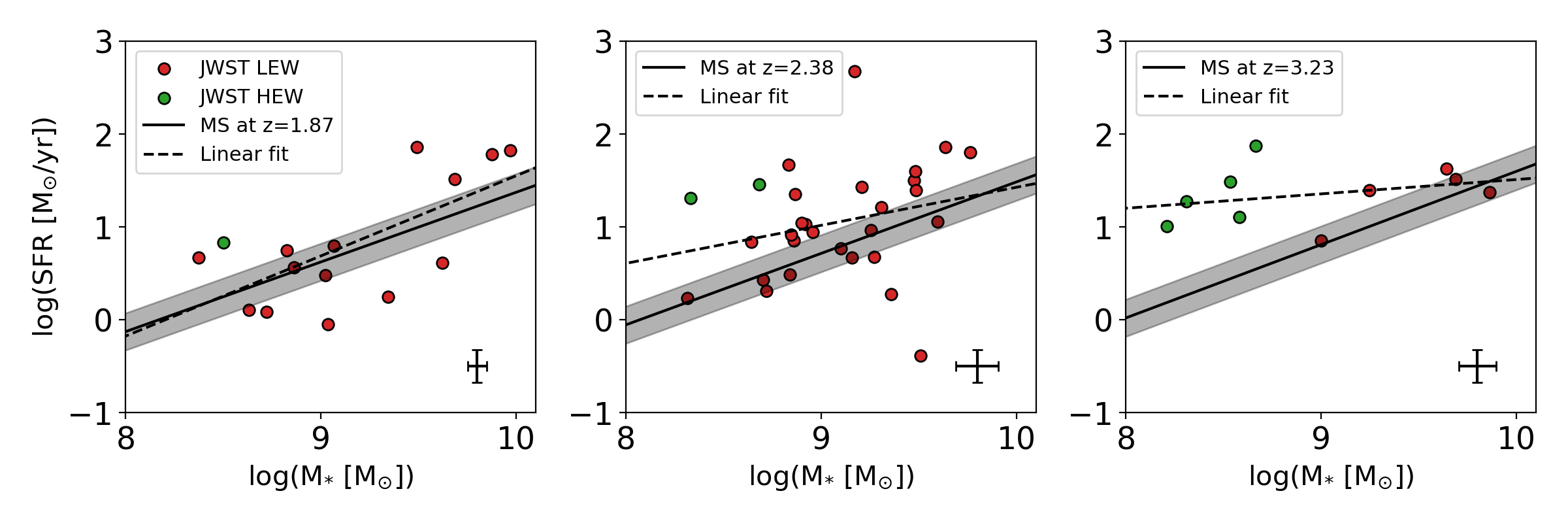}
        \caption{Star-forming main sequence for the JWST galaxies divided in three redshift bins, from left to right: $z$ $<$ 2, 2 $<$ $z$ $<$ 3 and $z$ $>$ 3. The red and green points represent the LEW and HEW galaxies, respectively. For each redshift bin, we show the SFMS from \cite{speagle14} considering the average redshift of the galaxies within the bin (black solid line) and the corresponding 1$\sigma$ scatter in the relation (black shaded region) and the linear fit to the galaxies in our sample (black dashed line).}
        \label{fig:SFMS}
    \end{figure*}

In Fig. \ref{fig:SFMS} we show the JWST galaxies in the SFR--M$_{*}$ diagram divided in three redshift intervals: $z$ $<$ 2, 2 $<$ $z$ $<$ 3 and $z$ $>$ 3. For each redshift interval, we show the linear fit to the data and also the SFMS from \cite{speagle14}, calculated at the average redshift of the galaxies within each bin, and the corresponding 1$\sigma$ scatter. Additionally, we colour code the LEW and HEW galaxies in red and green, respectively.

In all redshift intervals the HEW galaxies are located above the SFMS, but there are also LEW galaxies above the SFMS. Furthermore, the EW(H$\alpha$) of these LEW galaxies is not necessarily close to the threshold limit. This means that for a galaxy to have a significant nebular continuum contribution is not a sufficient condition to be forming stars at a rate above the typical value for its redshifts. We also see that for the $z$ $<$ 2 interval, the linear fit to our data is compatible with the SFMS in terms of slope and normalisation, being contained within the 1$\sigma$ scatter. However, moving to the higher redshift intervals, the linear fit to our data becomes increasingly flatter (becoming almost horizontal for the $z$ $>$ 3 interval) and inconsistent with the SFMS. This indicates that the impact of our quality criteria on the final sample is similar to that of a flux-limit selection.

To further refine the comparison between the physical properties of the JWST and SDSS samples, we identify analogous HEW galaxies in both datasets following two different criteria: 1) position on the X$_{\text{neb}}$--EW(H$\alpha$) relation and 2) position on the SFMS. For every HEW galaxy in the JWST sample, we identify HEW galaxies in the SDSS sample whose properties fall within a 0.2 dex distance of the corresponding JWST values according to each criteria. For the X$_{\text{neb}}$--EW(H$\alpha$) comparison, an average of 132 SDSS analogues was obtained, whereas for the SFMS-based comparison, the average number of analogues was 14. In the end, we compare the average value of relevant physical properties between the reference JWST galaxy and the selected SDSS analogues.

In Table \ref{tab:JWST_vs_SDSS_analogues}, we show the results of the comparison between the physical properties of HEW JWST galaxies and the selected SDSS analogues. For the physical properties considered, we computed the difference between each HEW JWST galaxy and the mean of its corresponding group of SDSS analogues. The mean of these differences for each property is the value we consider for this analysis.

    \begin{table}[!ht]
    \caption{Comparison of different physical properties between HEW analogue galaxies between the JWST and SDSS samples considering a X$_{\text{neb}}$--EW(H$\alpha$) and SFMS based selection.}
    \centering
        \begin{tabular}{ccc}
        \hline \hline
        Selection   & Property  & Avg. Difference [dex] \\ \hline
        \multirow{6}{*}{\begin{tabular}[c]{@{}c@{}}X$_{\text{neb}}$\\and\\EW(H$\alpha$)\end{tabular}}
         & M$_{*}$       & -0.04                 \\
         & t$_{L}$       & -0.40                 \\
         & Z$_{L}$       & -0.21                 \\
         & SFR           & 0.41                  \\
         & sSFR          & 0.46                  \\
         & A(H$\alpha$)  & -0.13                 \\ \hline
        \multirow{5}{*}{\begin{tabular}[c]{@{}c@{}}SFMS\end{tabular}}
         & t$_{L}$       & 0.23                  \\
         & Z$_{L}$       & -0.23                 \\
         & A(H$\alpha$)  & -0.28                 \\
         & X$_{\text{neb}}$     & -0.05                 \\
         & EW(H$\alpha$) & -0.07                 \\ \hline        
        \end{tabular}
    \tablefoot{We present the average difference, in dex, for each physical property between the JWST reference galaxies and the mean value of the identified SDSS analogues according to each criterion. A positive average difference means that JWST tends to obtain larger estimates relative to SDSS, while negative values mean the opposite.}
    \label{tab:JWST_vs_SDSS_analogues}
    \end{table}

The analogue selection based on the position in the X$_{\text{neb}}$--EW(H$\alpha$) relation, selects galaxies with similar stellar masses. However, the JWST galaxies exhibit higher SFR and sSFR, while the stellar age, metallicity, and dust extinction are lower. Considering the analogue selection based on the SFMS position, leads to galaxies with comparable EW(H$\alpha$) and by extension X$_{\text{neb}}$. On the other hand, the stellar age is on average higher and the stellar metallicity and dust extinction lower for the JWST galaxies. This is consistent with the results derived from Fig. \ref{fig:Prop_Abv_Bel_Thresh_Gal}.

\begin{figure}[!ht]
        \centering
        \includegraphics[scale=0.35]{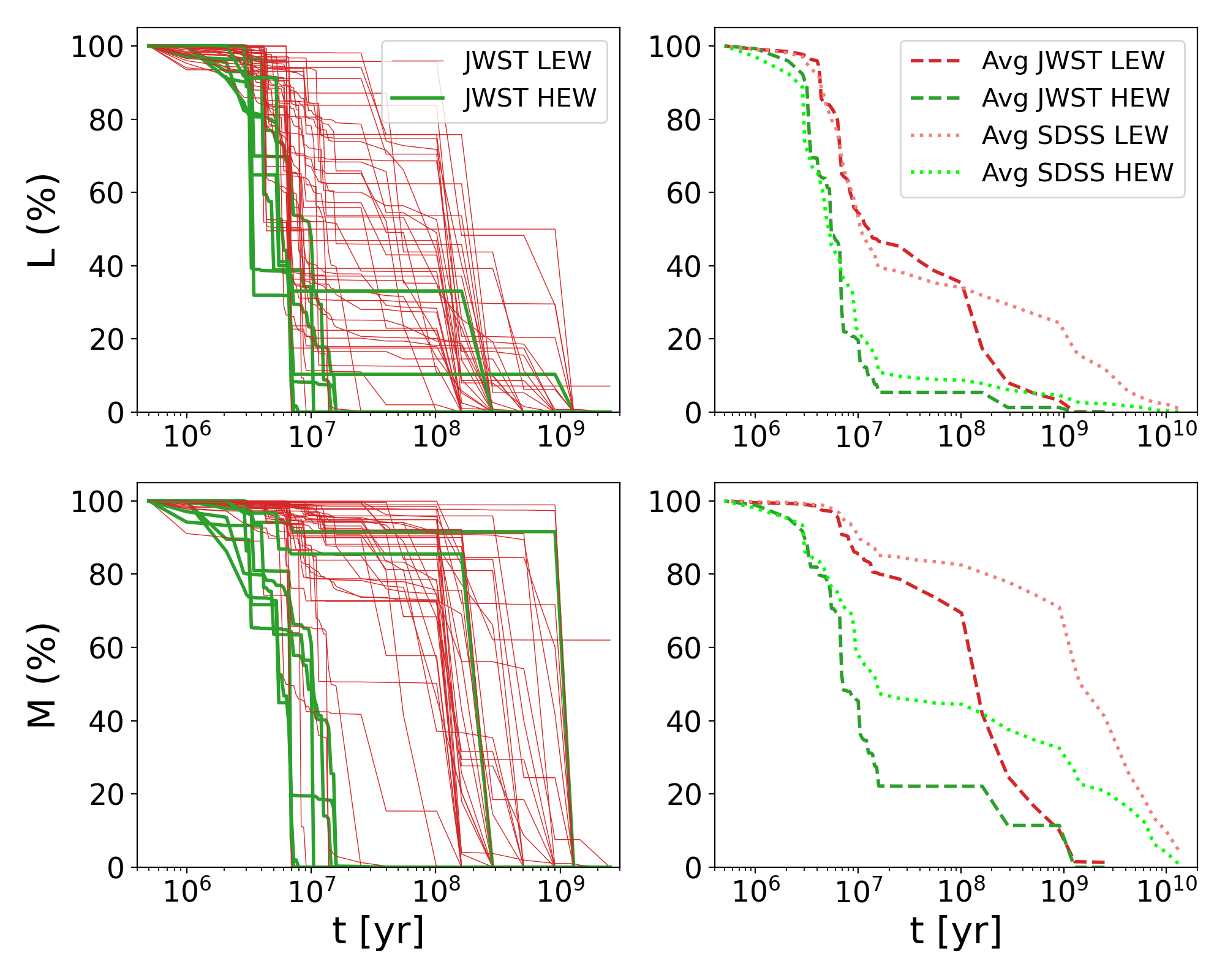}
        \caption{Comparison between the SFH of LEW (red) and HEW (green) galaxies for the JWST and SDSS samples. \textit{Left panels}: Cumulative light (top) and mass (bottom) contribution of the SSPs fitted by FADO to the JWST sample. \textit{Right panels}: Average cumulative light (top) and mass (bottom) contribution for the JWST (dashed lines) and SDSS (dotted lines) samples.}
        \label{fig:SFH_Abv_Bel_Thresh_Gal}
    \end{figure}

In Fig. \ref{fig:SFH_Abv_Bel_Thresh_Gal}, the left panels show the cumulative light (calculated at $\lambda$=4020 \AA) and mass contribution of the SSPs used by FADO to fit the JWST galaxies, with LEW and HEW galaxies represented in red and green, respectively. The right panels show the average cumulative light and mass contributions for the JWST (dashed lines) and SDSS (dotted lines) samples, divided in the LEW and HEW subsamples, in green and red colours, respectively. These can be seen as indicative of the star formation history (SFH) of the galaxies.

From the 8 HEW galaxies, 6 of them (75\%) only have light and mass contributions from SSPs younger than 20 Myr. Contrarily, for the LEW galaxies, 5 out of 46  (11\%) verify this condition. Expanding on this point, we find that from the 8 HEW galaxies, 7 (88\%) and 6 (75\%) have more than 80\% light and mass contribution, respectively, from the SSPs younger than 20 Myr. In contrast, from the 46 LEW galaxies, only 10 (22\%) and 5 (11\%) meet this condition. This shows that in the vast majority of cases the HEW galaxies are overwhelmingly dominated by extremely young SSPs, whereas for the LEW galaxies these SSPs have a less pronounced contribution.

Performing the same analysis to the SDSS data, for the LEW galaxies we obtain that 76 (31\%) and 25 (10\%) galaxies have more than 80\% light and mass contribution, respectively, from the SSPs younger than 20 Myr. For the HEW galaxies, 209 (81\%) and 115 (45\%) also satisfy this condition. Unlike what was obtained for the JWST sample, for the HEW galaxies, there is also a reasonable number of cases (98, corresponding to 38\%) for which the SSPs younger than 20 Myr have a mass contribution below 20\%.

This picture is further supported by analysing the average light and mass contribution for the JWST and SDSS samples. An immediate clear difference between the JWST and SDSS samples is the age of the oldest SSPs, around 1 Gyr and 10 Gyr for the JWST and SDSS samples, respectively. This difference is explained by the different redshift ranges covered by the two samples, which we translated into the set of SSP ages included in the spectral basis. Apart from this, the LEW galaxies have a comparable average light and mass contribution between JWST and SDSS. The average light contribution is also similar for the HEW galaxies between the JWST and SDSS samples, with SSPs younger than 20 Myr contributing to more than 90\% of the total luminosity. However, regarding the average mass contribution, while in the JWST sample SSPs younger than 20 Myr account for roughly 80\% of the total mass, in the SDSS sample these SSPs represent about 50\%, thus highlighting the existence of an underlying older stellar population.

\section{Discussion}
\label{sec:discussion}

One of the objectives of this work was to estimate the nebular contribution of high-redshift galaxies and relate it to previously established tracers and physical properties. In Sect. \ref{sec:FitQuality_NebCont} we showed that there is consistency between our results and the key findings from \cite{miranda25}: 1) strong linear correlation between X$_{\text{neb}}$ and the EW of H$\alpha$ and H$\beta$, 2) galaxies with high X$_{\text{neb}}$ values tend to be less massive and younger, but no clear correlation with stellar metallicity, and 3) for SF galaxies with X$_{\text{neb}}$ $\geq$ 8\% (EW(H$\alpha$) $\geq$ 500 \AA) the nebular continuum emission is significant and must be properly taken into account to obtain reliable physical properties estimates.

On the other hand, in our JWST sample there are no galaxies with EW(H$\alpha$) $>$ 1000 \AA. Including these objects in the analysis would enable the detailed analysis of the physical and evolutionary properties of such extreme objects at high redshifts, while also extending the sampled parameter space to the very high end of X$_{\text{neb}}$. Additionally, because the sample is not representative of the galaxy population at the covered redshifts, we cannot evaluate the predictions made in \cite{miranda25} about the expected fraction of galaxies with significant nebular contribution with increasing redshift.

The other objective of this work was to characterise the physical and evolutionary properties of galaxies with significant nebular continuum contribution. The analysis conducted in Sect. \ref{sec:HEW_galaxies_analysis} showed that these galaxies tend to be less massive, younger, with higher sSFR and smaller amounts of dust extinction relative to the galaxies without significant nebular contribution. It is relevant to note that this characterisation overlaps with the known properties of EELGs at both low and high redshift \citep[e.g.][]{cardamone09,vanderwel11,amorin15,brunker20,llerena24}.

Analysing this point in more detail, for the HEW galaxies we obtain a median stellar mass, light-weighted age and light-weighted metallicity of 10$^{8.52}$ M$_{\odot}$, 10$^{7.08}$ yr and 0.3 Z$_{\odot}$, respectively. For the SFR and sSFR, we obtain 10$^{1.29}$ M$_{\odot}$yr$^{-1}$ and 10$^{-7.13}$ yr$^{-1}$, respectively. Recent works studying EELGs with JWST data \citep[e.g.][]{boyett24,llerena24,llerena26,daikuhara25} have reported mean values and distributions that are broadly compatible with these estimates (considering that in these works a variety of different tools, methods and both photometric and spectroscopic data are used to obtain the estimates).

However, there are relevant differences between our results and the literature for the EW of H$\alpha$ and H$\beta$. While in our sample the maximum H$\alpha$ and H$\beta$ values are $\sim$1000 \AA\ and $\sim$300 \AA, respectively, in the literature we can find value that go as high as 3000 \AA\ for H$\alpha$ \citep[e.g][]{llerena24} and 600 \AA\ for H$\beta$ \citep[e.g.][]{boyett24}. This shows that there is a subgroup of extremely strong emitters that are not included in our sample, but that have already been observed in other works. Additionally, several works have analysed EELGs at $z$ $\sim$ 4--9, probing a relatively younger Universe than our sample. Both these discrepancies are a product of our sample selection criteria, as we have discussed throughout the text.

When comparing the JWST and SDSS analogue galaxies in terms of their X$_{\text{neb}}$ and EW(H$\alpha$) values, we see that the JWST galaxies tend to have comparable stellar mass, higher SF activity and lower stellar age, metallicity and dust extinction relative to the SDSS galaxies. Considering galaxy analogues following the position on the SFMS, the JWST galaxies have in general similar X$_{\text{neb}}$ and EW(H$\alpha$), higher stellar ages and lower stellar metallicity and dust extinction.

With respect to the evolutionary properties, the main difference between HEW and LEW galaxies is the significant light and mass contribution of SSPs younger than 20 Myr (see Fig. \ref{fig:SFH_Abv_Bel_Thresh_Gal}). We see that on average these SSPs contribute to around 90\% and 80\% of the total light and mass of HEW galaxies, respectively. In contrast, the contribution decreases to around 50\% and 20\% of the total light and mass of LEW galaxies, respectively. 

When comparing HEW galaxies between JWST and SDSS, our results show that there is a similar light contribution by the SSPs younger than 20 Myr ($\approx$90\%). However, the mass contribution of these SSPs is around 80\% for the JWST galaxies and 50\% for the SDSS galaxies. This means that the HEW galaxies are undergoing an intense star formation burst which contributes similarly to the total observed light for both JWST and SDSS. However, for the SDSS sample there is a larger amount of mass already formed in the past, thus leading to a lower mass contribution of younger SSPs. Nevertheless, in both cases a significant amount of the total mass of the galaxy was formed at recent times.

In summary, there is a clear difference between the evolutionary history of LEW and HEW galaxies, particularly the importance of the SSPs younger than 20 Myr. The comparison illustrates that while HEW galaxies at low and high redshifts share some characteristics, there are differences that point to distinct evolutionary contexts between the two. The physical significance of these results is worth exploring, particularly two aspects: 1) the difference in the relevance of the SSPs younger than 20 Myr between LEW and HEW galaxies and 2) the estimated extremely young ages of the HEW galaxies.

Relative to the first point, \cite{pappalardo21} shows the evolution of EW(H$\alpha$) over time assuming a continuous star formation model (see their Fig. 3). We see that following the initial burst of star formation, the EW(H$\alpha$) gradually decreases over time, reaching the EW(H$\alpha$)=500 \AA\ threshold after roughly 20 Myr, thus justifying the relevance of these SSPs to explain high EW(H$\alpha$) values. Regarding the interpretation of the high contribution of these SSPs to the total light and mass of the galaxy, we need to consider the definition of EW: the ratio of the line flux to the continuum level. For SF galaxies and focussing on the H$\alpha$ emission line, this implies that it also traces the connection between the current SFR (indicated by the line flux) and the average SFR over the lifetime of the galaxy (approximately captured by the continuum level). In this way, it is fundamental that the contribution to the total light and mass of the galaxy from SSPs younger than 20 Myr is sufficiently high so that the EW(H$\alpha$) reaches elevated values. In summary, a galaxy will have a significant nebular contribution only if it hosts young stellar populations (t$_{*}$ $<$ 20 Myr) capable of efficiently ionising the surrounding gas, and in sufficient numbers relative to older stellar populations such that their influence on the total light and mass is predominant.

Concerning the second point, for most HEW galaxies the contributions to the total light and mass are dominated by the SSPs younger than 20 Myr, thus leading to an average estimated age slightly above 10$^{7}$ yr. These values are low and it is difficult to conceive the absence of older, very faint stellar populations. It is likely that these stellar populations exist in the galaxy, but due to their high mass-to-light ratio they have a negligible contribution to the observed emission when compared to the young stellar populations, which have very low mass-to-light ratios. In the end, the optical spectrum is overwhelmingly dominated by young and massive stars and there are no discernible features arising from older stellar populations. This is the well-known outshining effect, apparent from numerous surface photometry studies of blue compact dwarf galaxies \citep[e.g.][]{loose86,kunth88,papaderos96,gomes16}, and has been shown to impact the estimated stellar masses up to one order of magnitude \citep[e.g.][]{papovich01,conroy13,sorba18,topping22,narayanan24}. Considering the impact of the outshining effect, it is likely that our stellar mass estimates are lower limits of the actual values, particularly for the HEW galaxies. Consequently, the estimated mass contributions of the SSPs younger than 20 Myr are likely an overestimation of the real values. However, we do not quantify the extent to which our estimates are influenced by the outshining effect.

As discussed in Sect. \ref{sec:sample_methods}, our sample is not representative of galaxy population at the covered redshifts, but rather a selection of galaxies with observed spectra that meet our quality criteria. This is the main caveat regarding the considered sample that should be addressed in future works to enable a generalisation of our findings. One way of addressing this caveat, would be to extend this study to include the low-resolution spectra from NIRSpec.The larger number of available low-resolution spectra (relatively to the medium-resolution ones), together with their higher continuum sensitivity, would both increase the sample size and allow us to probe a more diverse population of galaxies. Nevertheless, this work serves as a stepping stone towards a more complete study of HEW galaxies at high redshifts.

JWST already enables the observation of representative samples of galaxy population at redshifts similar to those of our sample and meeting the quality criteria considered in this work (e.g., the Blue Jay Survey, \citealt{belli25}). However, these samples have small sizes and observations of relatively rare objects, such as HEW galaxies, are insufficient to fully characterise this population. Future multi-object spectroscopic instruments, such as MOONS \citep[Multi-Object Optical and Near-infrared Spectrograph;][]{cirasuolo16,cirasuolo20}, with its extragalactic survey MOONRISE \citep{maiolino20}, will substantially increase the number of these objects, thus allowing the detailed study of galaxies with significant nebular continuum emission and help understand the relevance of these objects at Cosmic Noon.

\section{Summary and conclusions}
\label{sec:sum_and_conc}

Studying galaxies experiencing significant SF activity is essential, as they are a unique laboratory for a crucial phase of galaxy evolution. An important characteristic of these systems is their prominent nebular continuum emission in the optical and near-infrared regime. Accurately accounting for this emission and understanding its impact on the derived physical and evolutionary properties of galaxies requires high-quality spectroscopic data, which in the past were largely available only for galaxies in the local Universe. However, the groundbreaking JWST observations are now enabling the study of the nebular contribution of high-redshift galaxies.

We study a sample of 54 SF galaxies from DJA in the 1.7 $<$ $z$ $<$ 3.9 interval with the aim of analysing their nebular contribution. These galaxies have a broad coverage of the rest-frame optical wavelengths and main emission lines, plus sufficient S/N in the continuum to carry out full spectral fitting. We use the population spectral synthesis code FADO to fit the spectrum and estimate the stellar and nebular continua, from which we determine X$_{\text{neb}}$, as well as the physical and evolutionary properties of the galaxies.

We correlate the X$_{\text{neb}}$ estimates with the EW of H$\alpha$ and H$\beta$ and sSFR and compare them with the relations derived in \cite{miranda25} using SDSS data (see Fig. \ref{fig:Xneb_vs_tracers}). The results suggest that the relationship between X$_{\text{neb}}$ and the EW of H$\alpha$ and H$\beta$ are compatible between the JWST and SDSS data. Conversely, for the sSFR there is a slight offset between the two, with the same value of sSFR corresponding to higher X$_{\text{neb}}$ for the SDSS galaxies relative to those from JWST. However, the parameter space covered by the JWST and SDSS galaxies is compatible. We also test the Balmer break strength as a tracer of X$_{\text{neb}}$, finding a moderate negative correlation between the two.

We further examine how X$_{\text{neb}}$ relates to different physical properties (see Fig. \ref{fig:Xneb_vs_galprop}). We find that larger X$_{\text{neb}}$ values tend to correspond to lower stellar masses and younger ages. Although the correlation between X$_{\text{neb}}$ and stellar mass is comparable between JWST and SDSS, the correlation with stellar age is steeper for JWST relative to SDSS, which is attributed to the younger age of the Universe. For the stellar metallicity, no correlation with X$_{\text{neb}}$ was observed, similar to the results using SDSS data. 

We compare the physical properties estimates using FADO in full-consistency mode and in pure-stellar mode as a function of the EW(H$\alpha$) value (see Fig. \ref{fig:FD_FC_vs_PS}). The results show that the difference between the two modes becomes significant when EW(H$\alpha$) $\ge$ 500 \AA\ (equivalent to X$_{\text{neb}}$ $\ge$ 8\%), thus supporting the applicability at higher redshifts of the threshold derived in \cite{miranda25} using SDSS data. For these galaxies, the nebular continuum emission is so significant that properly accounting for it becomes fundamental to accurately retrieve the physical and evolutionary properties.

Next, we divide the JWST and SDSS samples into galaxies with and without significant nebular contribution (HEW and LEW galaxies, respectively), according to the previously established threshold and compare their physical and evolutionary properties (see Figs. \ref{fig:Prop_Abv_Bel_Thresh_Gal}, \ref{fig:SFMS} and \ref{fig:SFH_Abv_Bel_Thresh_Gal} and Table \ref{tab:JWST_vs_SDSS_analogues}). We find that for both JWST and SDSS, HEW galaxies tend to be less massive, younger, with higher SF activity and smaller amounts of dust extinction relative to LEW galaxies. Furthermore, HEW galaxies are systematically found above the SFMS. However, being above the SFMS alone does not guarantee a significant nebular contribution, since some LEW galaxies also occupy this region.

Looking into the evolutionary properties, we find that for a galaxy to have a significant nebular contribution, SSPs younger than 20 Myr must have a significant contribution to the overall light and mass of the galaxy. This result arises from the fact that young and massive stars are the ones responsible for actively ionising the nebular gas. However, due to the outshining effect, it is likely that the mass contributions of these young SSPs are overestimated in this work.

The selection criteria followed in this work are based on the quality of the spectra being sufficient to carry out spectral fitting. Due to this fact, the obtained sample is not representative of the galaxy population at the covered redshifts and hampers the generalisation of results and the verification of the predictions made in \cite{miranda25} regarding the fraction of galaxies with significant nebular contribution at higher redshifts. Despite this caveat, we achieved the overarching goal of estimating the nebular contribution of high-redshift galaxies and study their physical and evolutionary properties. In this context, the present work represents an initial step toward more comprehensive studies of nebular continuum emission in high-redshift galaxies based on larger and more representative spectroscopic samples.

\begin{acknowledgements}

This work was supported by Fundação para a Ciência e a Tecnologia (FCT) through national funds under the research grant UID/04434/2025 (DOI 10.54499/UID/04434/2025). H.M. acknowledges support by FCT through the PhD Fellowship 2022.12891.BD. C.P. work is funded by national funds through FCT, I.P., under the support of the FCT Tenure 1st Edition (position reference 2023.15441.TENURE.029). P.P. acknowledges support by FCT through the grant 2024.17990.PEX, and Principal Investigator contract CIAAUP-092023-CTTI. Some of the data products presented herein were retrieved from the Dawn JWST Archive (DJA). DJA is an initiative of the Cosmic Dawn Center (DAWN), which is funded by the Danish National Research Foundation under grant DNRF140. Funding for the SDSS and SDSS-II has been provided by the Alfred P. Sloan Foundation, the Participating Institutions, the National Science Foundation, the U.S. Department of Energy, the National Aeronautics and Space Administration, the Japanese Monbukagakusho, the Max Planck Society, and the Higher Education Funding Council for England. The SDSS Web Site is http://www.sdss.org/. The SDSS is managed by the Astrophysical Research Consortium for the Participating Institutions. The Participating Institutions are the American Museum of Natural History, Astrophysical Institute Potsdam, University of Basel, University of Cambridge, Case Western Reserve University, University of Chicago, Drexel University, Fermilab, the Institute for Advanced Study, the Japan Participation Group, Johns Hopkins University, the Joint Institute for Nuclear Astrophysics, the Kavli Institute for Particle Astrophysics and Cosmology, the Korean Scientist Group, the Chinese Academy of Sciences (LAMOST), Los Alamos National Laboratory, the Max-Planck-Institute for Astronomy (MPIA), the Max-Planck-Institute for Astrophysics (MPA), New Mexico State University, Ohio State University, University of Pittsburgh, University of Portsmouth, Princeton University, the United States Naval Observatory, and the University of Washington.

\end{acknowledgements}

\bibliographystyle{aa} 
\bibliography{biblio}

\begin{appendix}

\onecolumn
\section{Example fits with FADO}
\label{appendix_FADO_Example}

\begin{figure}[!ht]
        \sidecaption
        \centering
        \includegraphics[scale=0.35]{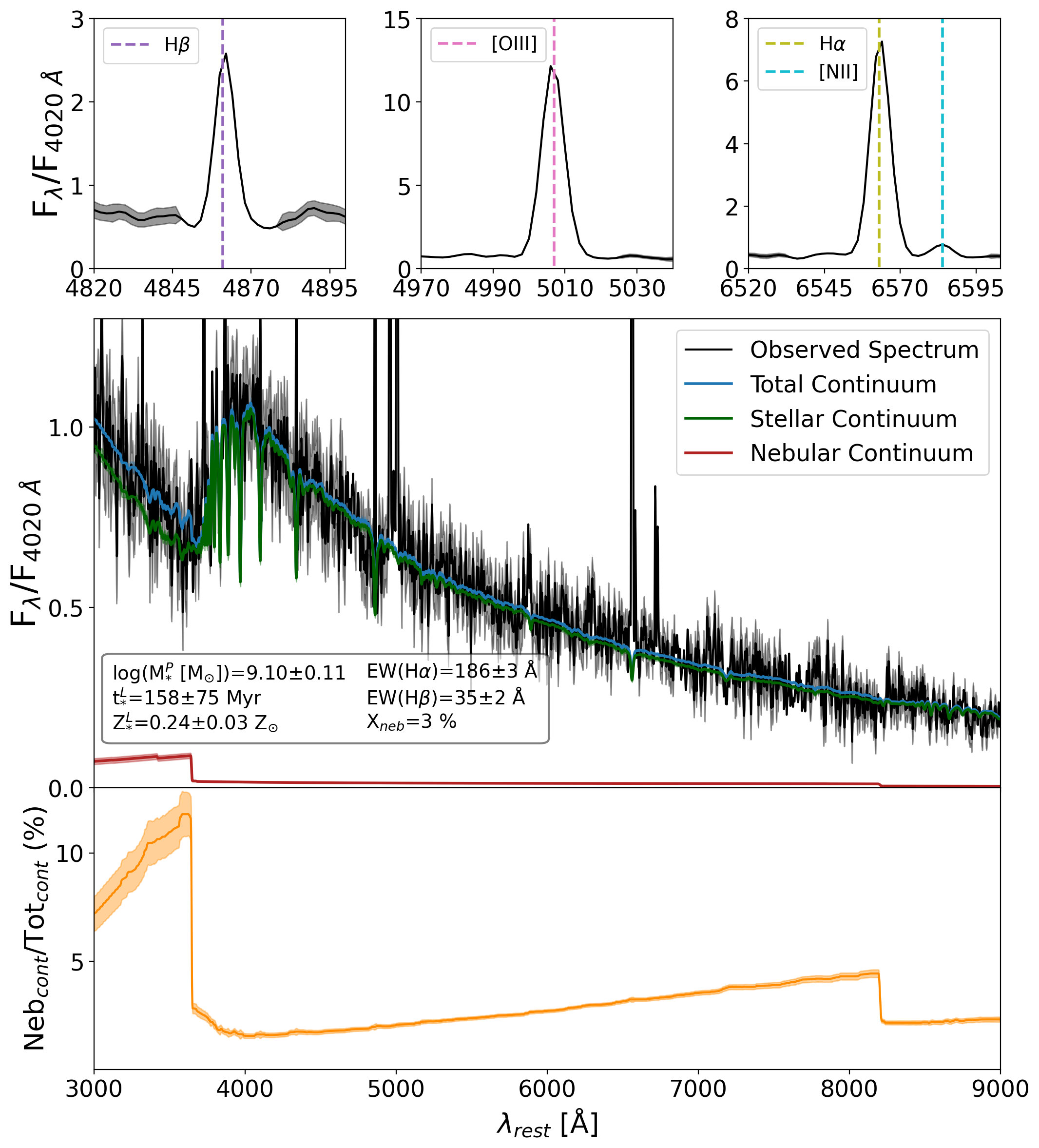}
        \caption{Model fitted by FADO to galaxy with S/N $\simeq$ 7, representative of lowest value for our sample. Galaxy with DJA \texttt{uid} 29968 and $z$ $\simeq$ 2.83, observed by the AURORA survey \citep{shapley25}. \textit{Upper Panels:} Observed spectrum, plus zoom-in on BPT emission lines, (black line) and total, stellar and nebular continuum fitted by FADO (blue, green and red lines, respectively). The shaded regions represent the uncertainty in the observed spectrum and estimated models. The y-axis shows the normalised flux at $\lambda$ = 4020 \AA. The inset box shows FADO estimates for some properties and associated uncertainties: presently available stellar mass ($M_{*}^{P}$), light-weighted stellar age ($t_{*}^{L}$) and metallicity ($Z_{*}^{L}$), H$\alpha$ and H$\beta$ EW and nebular contribution (X$_{\text{neb}}$). \textit{Bottom Panel:} Ratio between the nebular and total continuum fitted by FADO as a function of wavelength, with the shaded region representing the estimated uncertainty.}
        \label{fig:Fit_Example_Low}
    \end{figure}

\begin{figure}[!ht]
        \sidecaption
        \centering
        \includegraphics[scale=0.35]{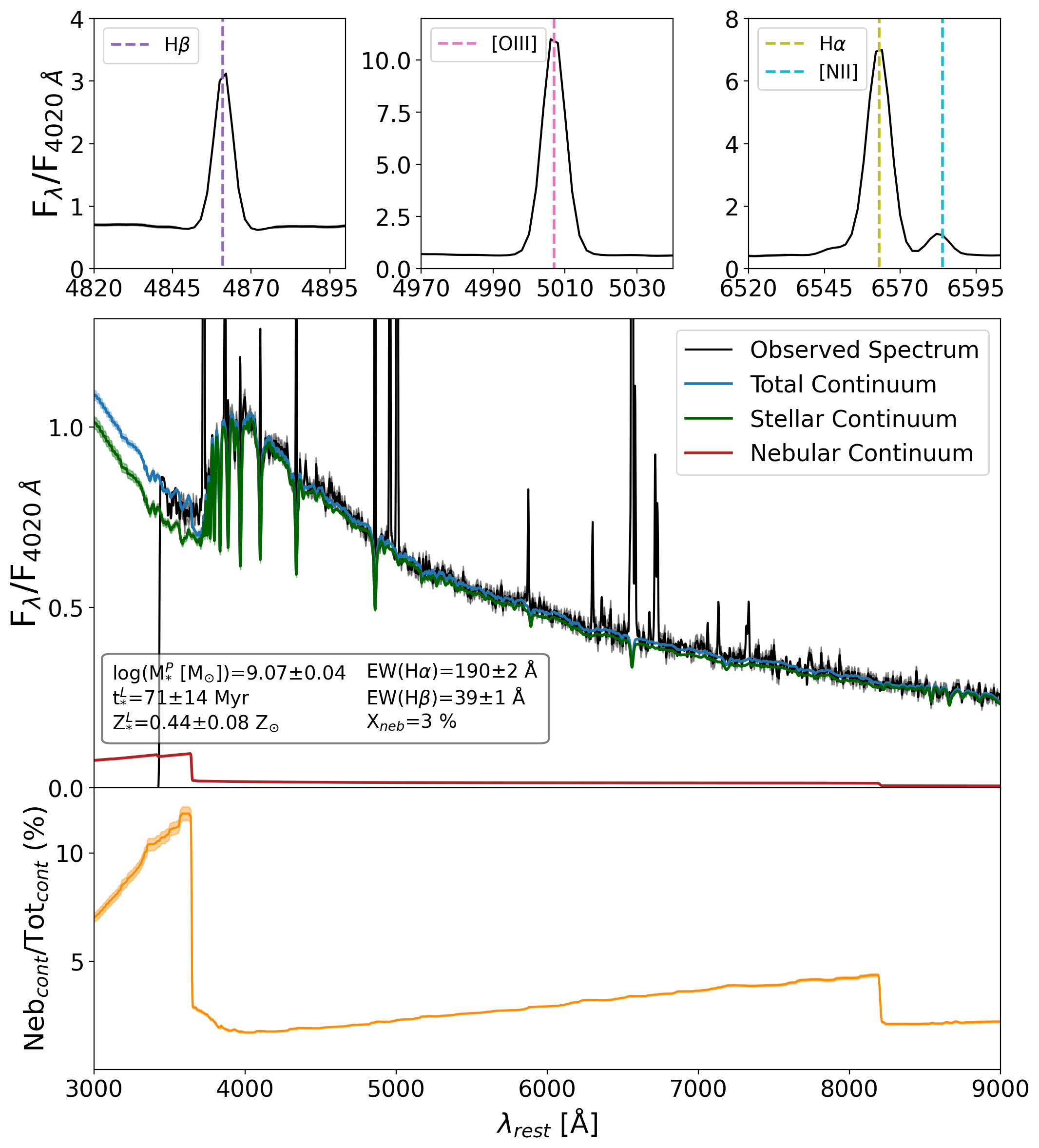}
        \caption{Same as Fig. \ref{fig:Fit_Example_Low}, but for galaxy with S/N $\simeq$ 30, representative of highest value for our sample. Galaxy with DJA \texttt{uid} 13815 and $z$ $\simeq$ 1.85, observed by the MARTA survey \citep{curti26}.}
        \label{fig:Fit_Example_High}
    \end{figure}

Figures \ref{fig:Fit_Example_Low} and \ref{fig:Fit_Example_High} show the FADO fit to representative galaxies from our sample with the lowest and highest mean S/N, respectively. The upper panels show the observed spectrum, plus a zoom-in on the BPT emission lines, and the fitted models by FADO with the respective uncertainties represented as shaded regions. The bottom panel shows the ratio between the fitted nebular and total continuum as a function of wavelength.

\begin{figure}[!ht]
        \sidecaption
        \centering
        \includegraphics[scale=0.35]{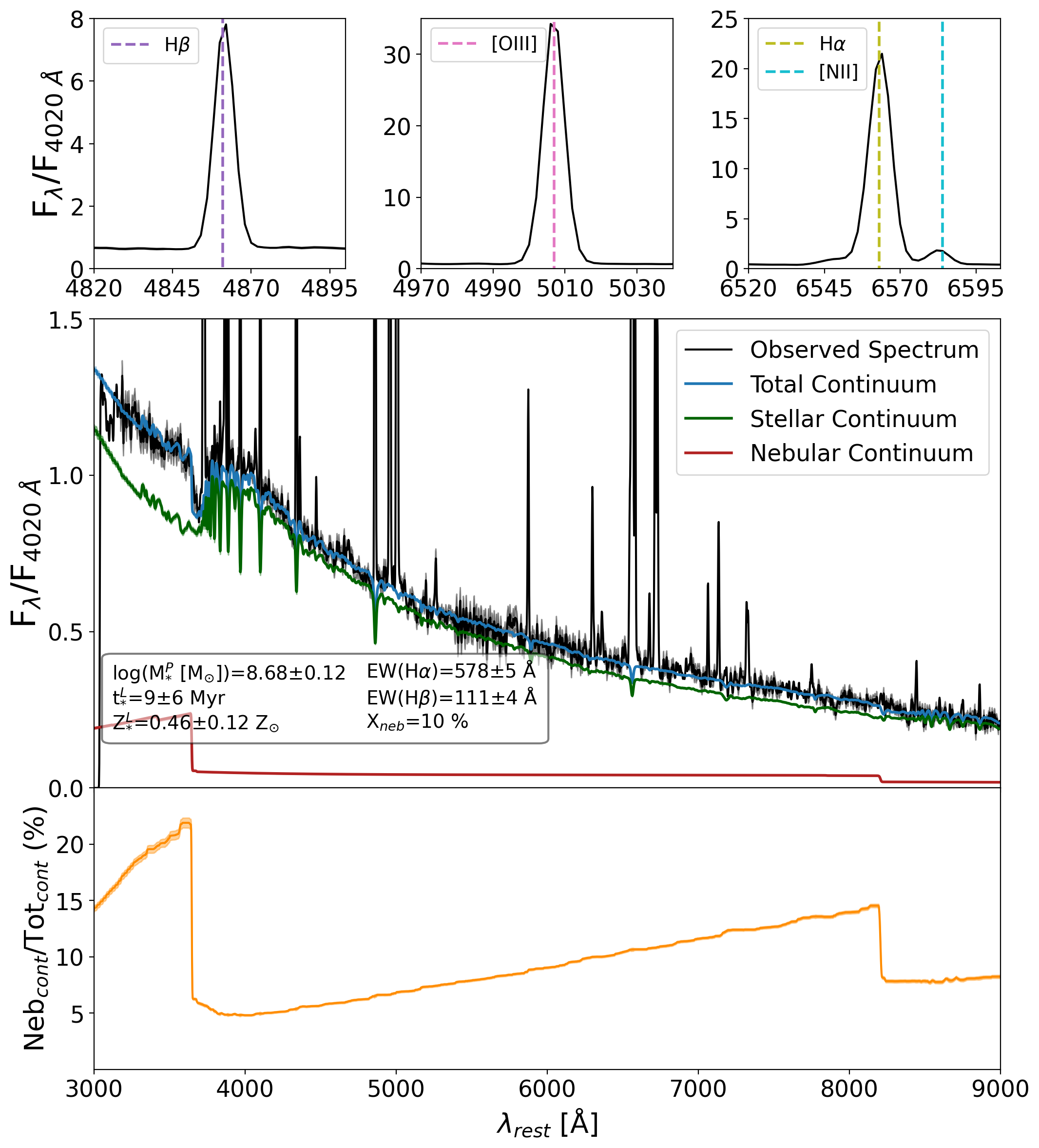}
        \caption{Same as Fig. \ref{fig:Fit_Example_Low}, but for galaxy with X$_{\text{neb}}$ $\simeq$ 10\%, representative of a galaxy at the threshold for significant nebular continuum contribution. Galaxy with DJA \texttt{uid} 13808 and $z$ $\simeq$ 2.22, observed by the MARTA survey \citep{curti26}.}
        \label{fig:Fit_Example_MidXneb}
    \end{figure}

\begin{figure}[!ht]
        \sidecaption
        \centering
        \includegraphics[scale=0.35]{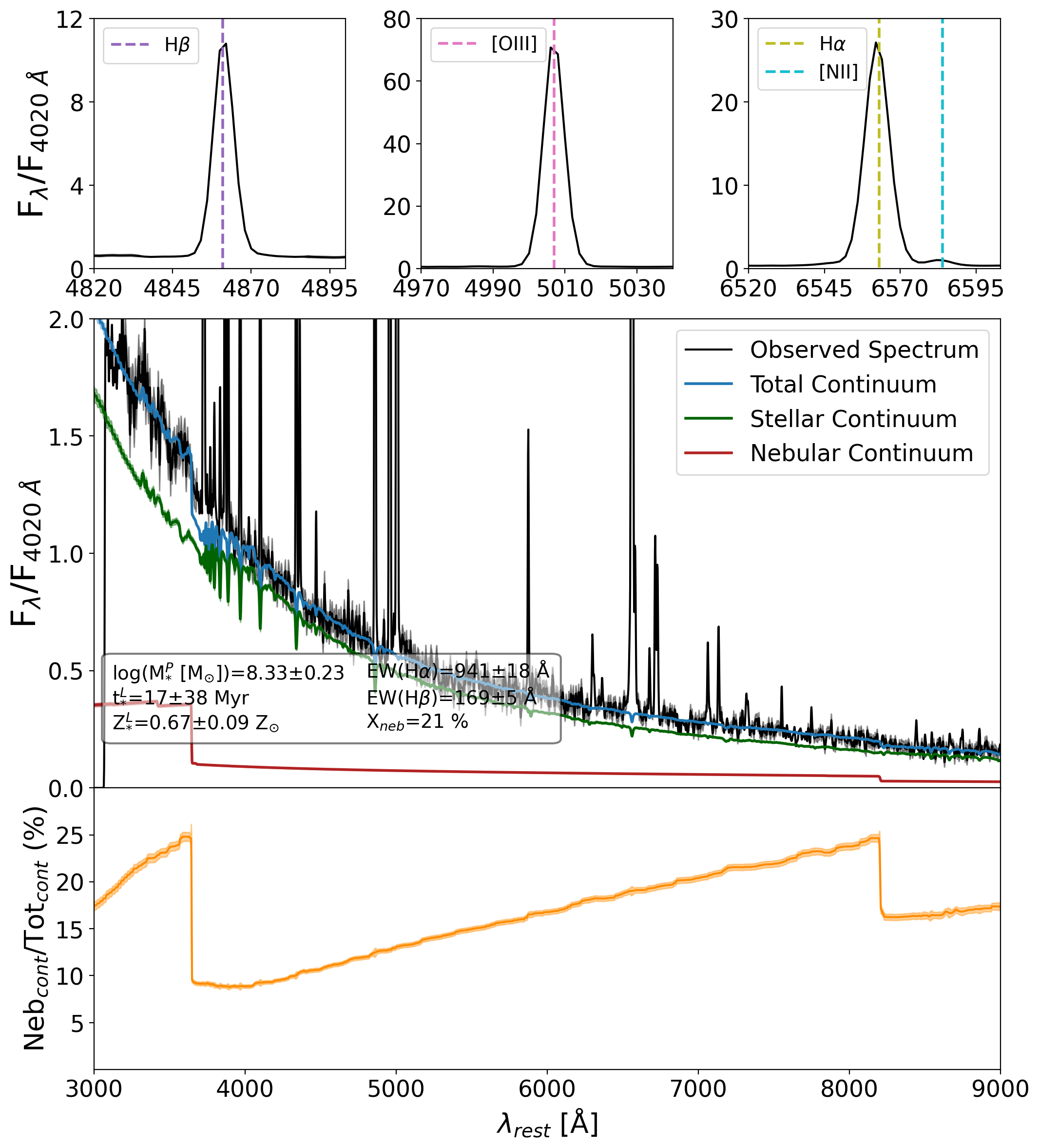}
        \caption{Same as Fig. \ref{fig:Fit_Example_Low}, but for galaxy with X$_{\text{neb}}$ $\simeq$ 21\%, representative of a galaxy with highest value for our sample. Galaxy with DJA \texttt{uid} 29931 and $z$ $\simeq$ 2.19, observed by the AURORA survey \citep{shapley25}.}
        \label{fig:Fit_Example_HighXneb}
    \end{figure}

Figures \ref{fig:Fit_Example_MidXneb} and \ref{fig:Fit_Example_HighXneb} show a FADO fit to representative galaxies from our sample with higher values of nebular continuum contribution, X$_{\text{neb}}$ = 10\% and X$_{\text{neb}}$ = 21\%, respectively. The figures follow the same structure as the previous ones.

\FloatBarrier
\clearpage

\section{Comparison of FADO estimates in full-consistency and pure-stellar mode}
\label{appendix_FADO_FCvsPS}

\begin{figure}[!ht]
        \centering
        \includegraphics[scale=0.40]{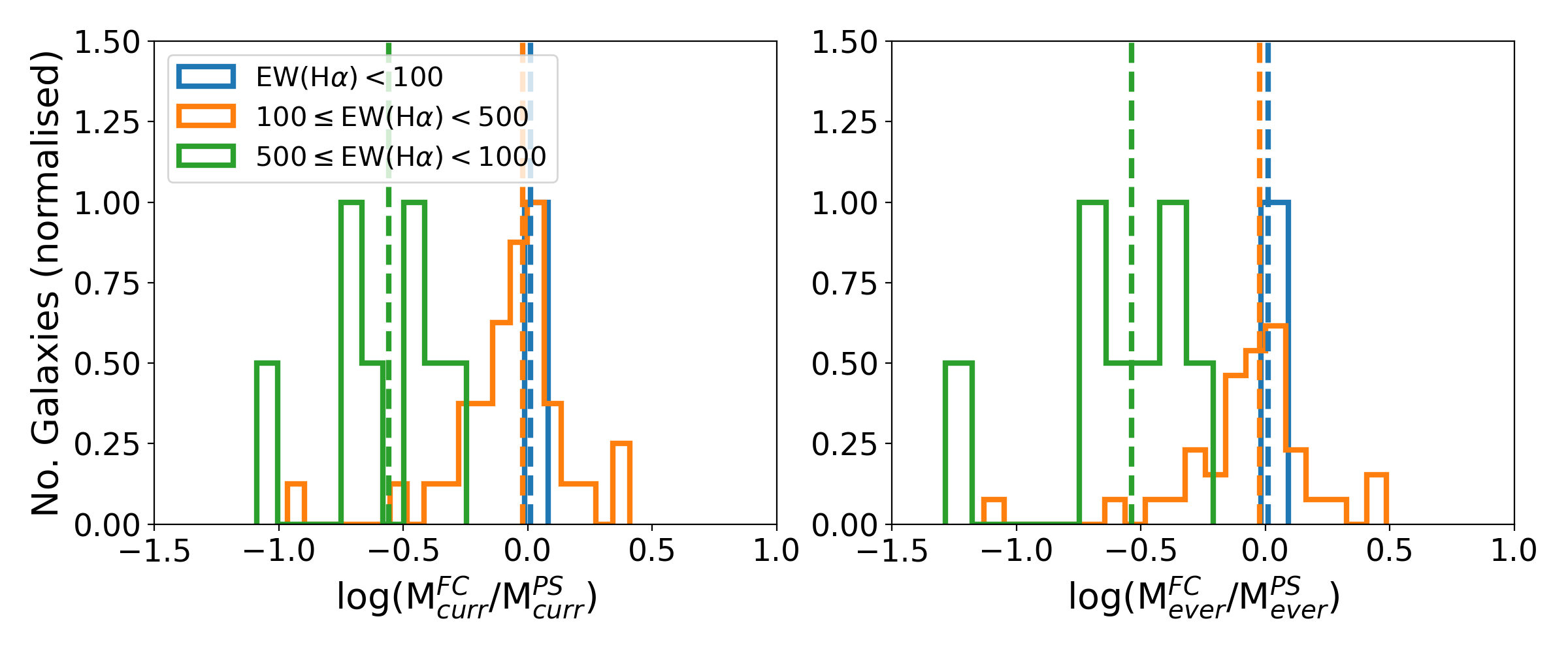}
        \caption{Distribution of the logarithmic difference between the currently available stellar mass (M$_{curr}$, left panel) and total ever formed stellar mass (M$_{ever}$, right panel) estimated by FADO in full-consistency (FC) and pure-stellar (PS) mode, for three EW(H$\alpha$) bins: EW(H$\alpha$) $<$ 100 \AA\ (blue), 100 $\leq$ EW(H$\alpha$) $<$ 500 \AA\ (orange) and 500 $\leq$ EW(H$\alpha$) $<$ 1000 \AA\ (green). The distributions are normalised so that the maximum value is equal to one. The vertical dashed lines represent the median value of each distribution.}
        \label{fig:FD_FC_vs_PS_M}
    \end{figure}

\begin{figure}[!ht]
        \centering
        \includegraphics[scale=0.40]{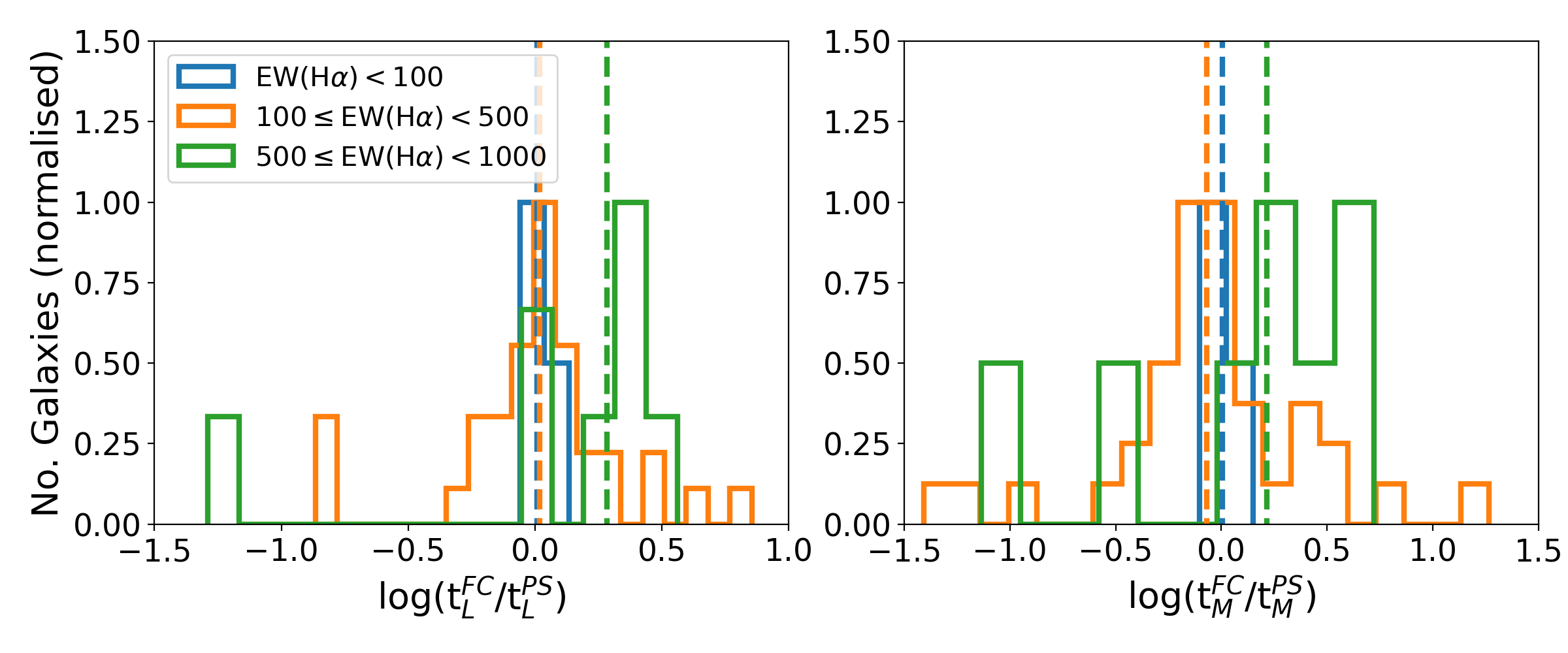}
        \caption{Same as Fig. \ref{fig:FD_FC_vs_PS_M}, but for the stellar age weighted by light (t$_{L}$, left panel) and by mass (t$_{M}$, right panel).}
        \label{fig:FD_FC_vs_PS_t}
    \end{figure}

\begin{figure}[!ht]
        \centering
        \includegraphics[scale=0.40]{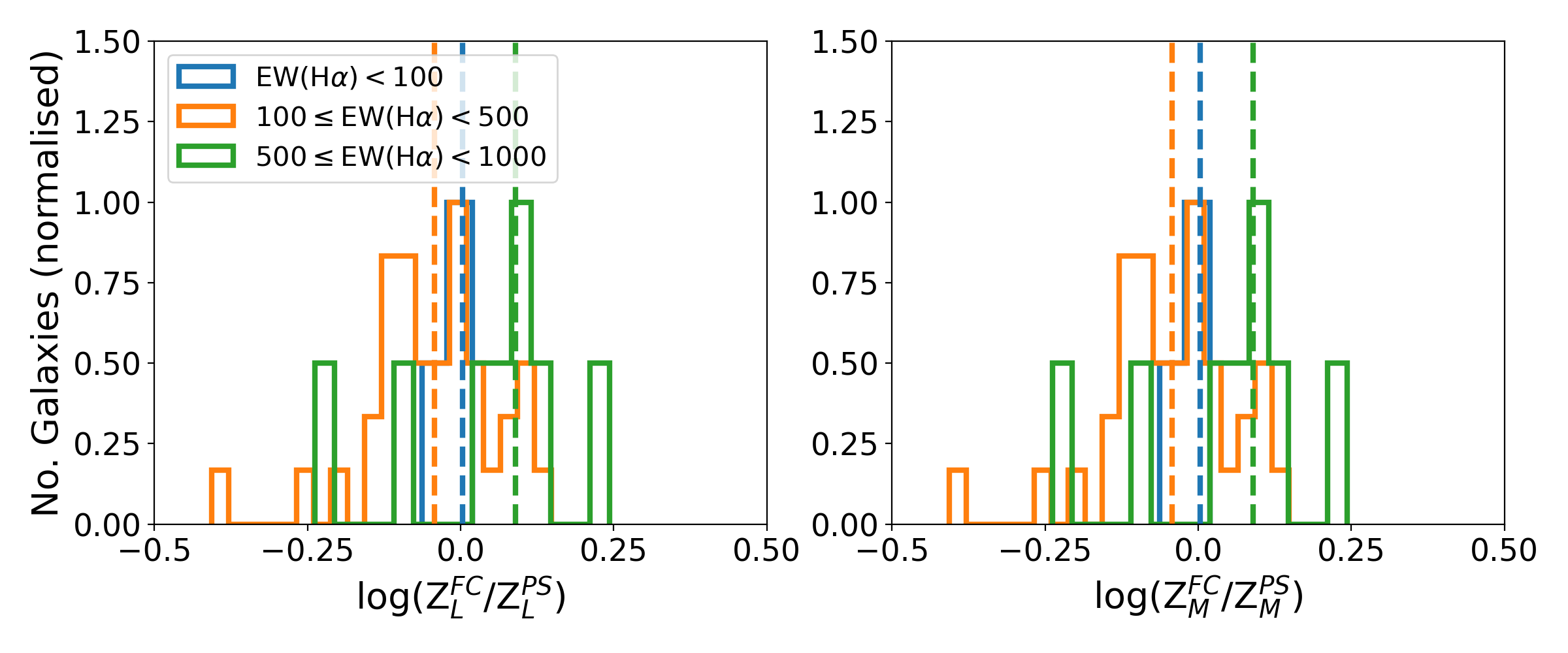}
        \caption{Same as Fig. \ref{fig:FD_FC_vs_PS_M}, but for the stellar metallicity weighted by light (Z$_{L}$, left panel) and by mass (Z$_{M}$, right panel).}
        \label{fig:FD_FC_vs_PS_Z}
    \end{figure}

Figures \ref{fig:FD_FC_vs_PS_M}, \ref{fig:FD_FC_vs_PS_t}, and \ref{fig:FD_FC_vs_PS_Z} show the distribution of the logarithm of the ratio between the estimates obtained by FADO in full-consistency (FC) and pure-stellar (PS) mode for the stellar mass, age and metallicity, respectively. In the plots, the positive values correspond to FC mode obtaining larger estimates relative to the PS mode and negative values correspond to the opposite case.

\clearpage

\begin{figure}[!ht]
        \sidecaption
        \centering
        \includegraphics[scale=0.3]{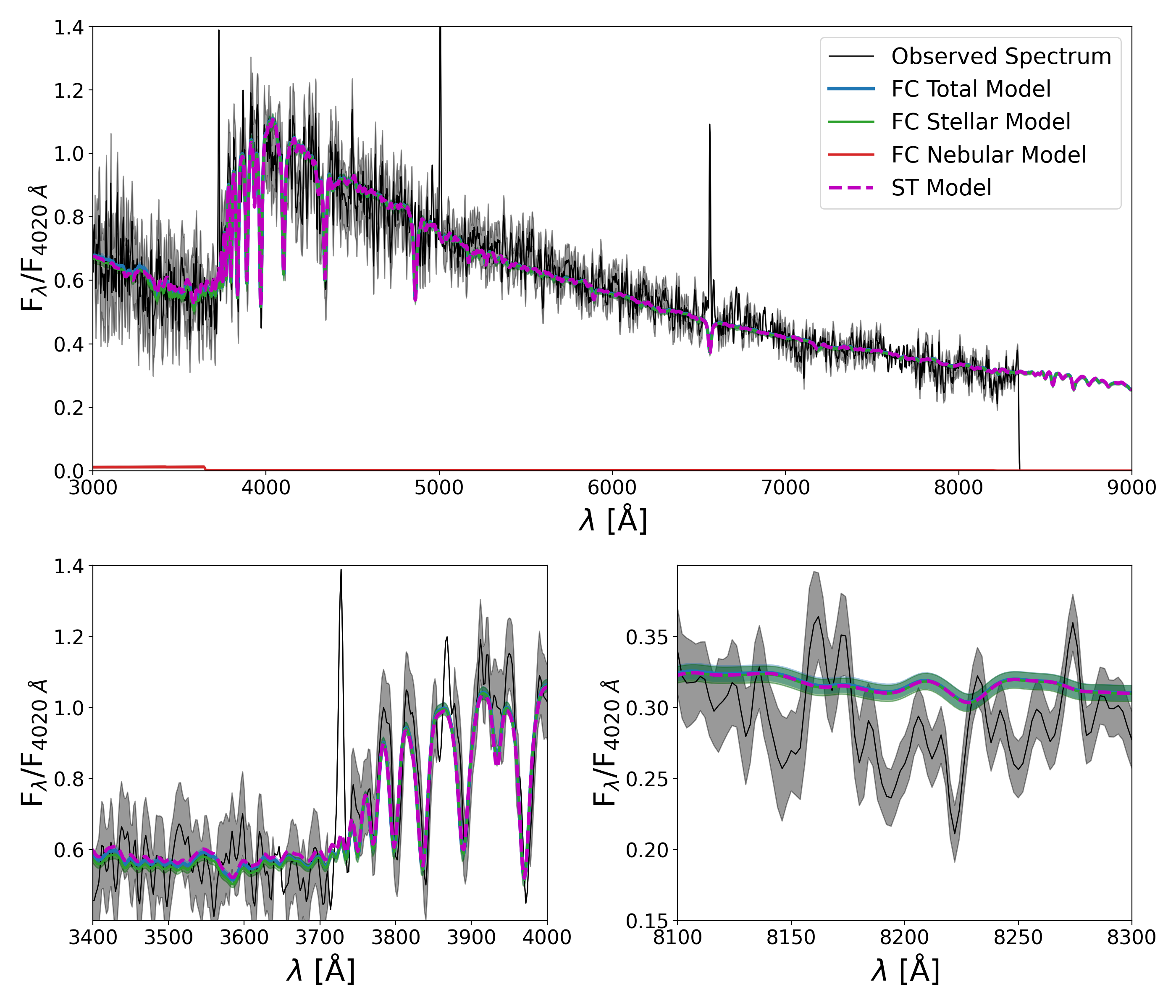}
        \caption{Comparison of the models fitted by FADO in FC and PS mode to galaxy with X$_{\text{neb}}$ $<$ 1\%. Galaxy with DJA \texttt{uid} 25861 and $z$ $\simeq$ 2.66, observed by the LyC22 survey \citep{schaerer21}. \textit{Upper Panels:} Observed spectrum (black line) and total, stellar and nebular continua fitted by FADO FC mode (blue, green and red lines, respectively), plus total continuum fitted by FADO PS mode (magenta line). The shaded regions represent the uncertainty in the observed spectrum and estimated models. The y-axis shows the normalised flux at $\lambda$ = 4020 \AA. \textit{Bottom Panels:} Zoom-in on the Balmer and Paschen discontinuities (left and right, respectively).}
        \label{fig:FD_FC_vs_PS_Example_LowXneb}
    \end{figure}

\begin{figure}[!ht]
        \centering
        \sidecaption
        \includegraphics[scale=0.3]{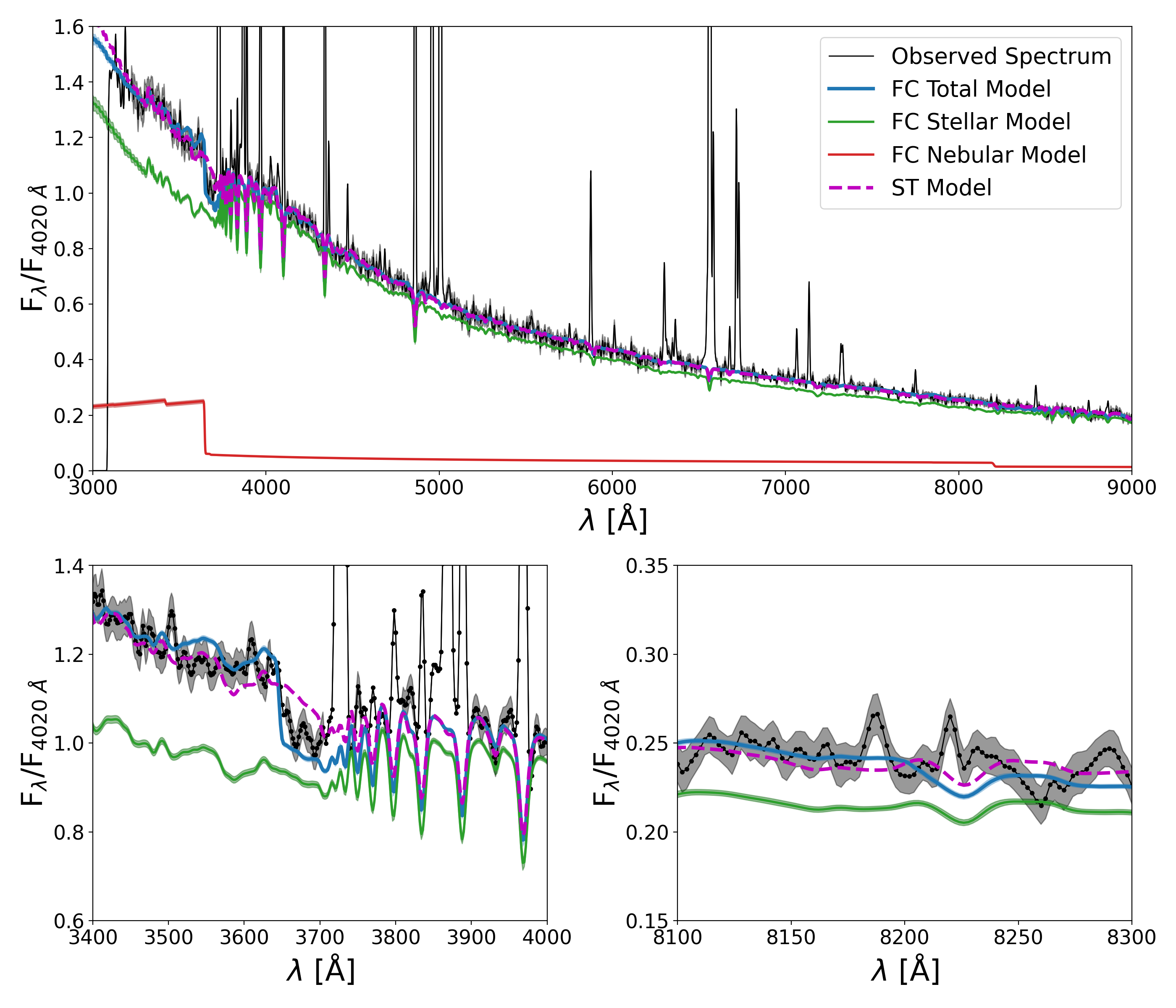}
        \caption{Same as Fig. \ref{fig:FD_FC_vs_PS_Example_LowXneb}, but for galaxy with X$_{\text{neb}}$ $\simeq$ 8\%. Galaxy with DJA \texttt{uid} 29934 and $z$ $\simeq$ 2.17, observed by the AURORA survey \citep{shapley25}}
        \label{fig:FD_FC_vs_PS_Example_HighXneb}
    \end{figure}

Figures \ref{fig:FD_FC_vs_PS_Example_LowXneb} and \ref{fig:FD_FC_vs_PS_Example_HighXneb} show the comparison between the fitted models by FADO when applied in FC and PS mode. The upper panel shows the full spectrum together with the different fitted continua, whereas the bottom panels show a zoom-in on the Balmer and Paschen discontinuities.

\end{appendix}

\end{document}